\documentclass{aa}

\usepackage{newtxtext,newtxmath}

\usepackage[T1]{fontenc}

\DeclareRobustCommand{\VAN}[3]{#2}
\let\VANthebibliography\thebibliography
\def\thebibliography{\DeclareRobustCommand{\VAN}[3]{##3}\VANthebibliography}

\usepackage[dvipsnames]{xcolor}
\usepackage{graphicx}	%
\usepackage{amsmath}	%
\usepackage[normalem]{ulem}
\usepackage{hyperref}

\usepackage{cleveref}

\newcommand{\Nyx}{\textsc{nyx}}
\newcommand{\lyanna}{\textsc{Ly$\alpha$NNA}}
\newcommand{\lya}{Ly$\alpha$}
\newcommand{\sansa}{\textsc{Sansa}}

\newcommand{\HI}{\ion{H}{i}}

\newcommand{\HeIII}{\ion{He}{iii}}

\begin{document}

\title{\lyanna: A Deep Learning Field-level Inference Machine for the Lyman-$\alpha$ Forest}

\author{
Parth Nayak \inst{1,2}\fnmsep\thanks{\email{parth.nayak@physik.lmu.de}} \and
Michael Walther \inst{1,2}\and
Daniel Gruen \inst{1,2}\and
Sreyas Adiraju \inst{3,1}
}
\institute{
University Observatory, Faculty of Physics, Ludwig-Maximilians-Universit\"at, Scheinerstr. 1, 81679 Munich, Germany \and
Excellence Cluster ORIGINS, Boltzmannstr. 2, 85748 Garching, Germany \and
Department of Psychology, Columbia University, New York, NY 10027, USA
}

\date{Received November 3, 2023; accepted May 29, 2024}

\label{firstpage}

\abstract{%
The inference of astrophysical and cosmological properties from the Lyman-$\alpha$ forest conventionally relies on summary statistics of the transmission field that carry useful but limited information.
We present a deep learning framework for inference from %
the Lyman-$\alpha$ forest at %
field-level. %
This framework consists of a 1D residual convolutional neural network (ResNet) that extracts spectral features and performs regression on thermal parameters of the IGM that characterize the power-law temperature-density relation. We train this supervised machinery using a large set of mock absorption spectra from %
{\Nyx} hydrodynamic simulations at $z=2.2$ with a range of thermal parameter combinations (labels). %
We employ Bayesian optimization %
to find an optimal set of hyperparameters for our network, and then employ a committee of 20 neural networks for increased statistical robustness of the network inference.   %
In addition to the parameter point predictions, our machine also provides a self-consistent estimate of their covariance matrix %
with which we construct a pipeline %
for inferring the posterior distribution of the parameters. 
We compare the results of our framework with the traditional summary (PDF and power spectrum of transmission) based approach in terms of the area of the 68\% credibility regions as our figure of merit (FoM). In our study of the information content of perfect (noise- and systematics-free) {\lya} forest spectral data-sets, we find a significant %
tightening of the posterior constraints --- factors of 10.92 and 3.30 in FoM over power spectrum only and jointly with PDF, respectively ---  %
that is the consequence of recovering the relevant parts of information that are not carried by the classical summary statistics.  %
}%

\maketitle

\section{Introduction} \label{sec:intro}

The characteristic arrangement of {\lya} absorption lines in the spectra of distant quasars, commonly known as the ``{\lya} forest'' \citep{Lynds:1971}, has been shown to be a unique probe of the physics of the Universe at play over a wide window of cosmic history ($z \lesssim 6$).
As the continua of %
emission by the quasars traverse the diffuse intergalactic gas, %
resonant scattering by the neutral Hydrogen leads to a net %
absorption of the radiation at the wavelength of the {\lya} transition \citep{Gunn_Peterson_1965ApJ...142.1633G}. %
In an expanding Universe where spectral redshift $z$ is a proxy of distance, 
a congregation of absorber clouds in the intergalactic medium (IGM) along a quasar sightline imprints a dense forest of {\lya} absorption lines on their spectra.
Due to cosmic reionization of Hydrogen being largely complete by $z \sim 6$ (e.g. \citealt{McGreer_Reion_2015MNRAS.447..499M}), its neutral fraction $x_{\HI}$ within the IGM is extremely minute, yet sufficient to produce this unique feature that enables a continuous trace of the cosmic gas. %

The observations of the {\lya} forest, through the advent of high-resolution instruments (e.g. Keck/HIRES and VLT/UVES) as well as large-scale structure surveys (e.g. eBOSS \citep{eBOSS_Main_2013AJ....145...10D} and DESI \citep{DESI_OVERVIEW}), %
have delivered a wealth of information about the nonlinear matter distribution on sub-Mpc scales, thermal properties of the intergalactic gas, and large scale structure.
Not only is the {\lya} forest an extremely useful tool to study the thermal evolution of the intergalactic medium (IGM) and reionization (as demonstrated, e.g., by %
\citealt{Becker_Inference_curvature_2011MNRAS.410.1096B},
\citealt{Walther_IGM_2019ApJ...872...13W}, \citealt{Boera_Inference_P1D_2019ApJ...872..101B}, \citealt{Gaikwad_IGM_2021MNRAS.506.4389G}), 
but it has also opened up avenues for constraining fundamental cosmic physics. Chief among those are baryonic acoustic oscillations (BAO) and cosmic expansion (e.g., \citealt{Slosar_BAO_2013JCAP...04..026S}, \citealt{Busca_BAO_2013A&A...552A..96B}, \citealt{eBOSS_BAO_Lya_2020ApJ...901..153D}, \citealt{Gordon_DESI-BAO_2023arXiv230810950G}, \citealt{Cuceu_Cosmo_2023PhRvL.130s1003C}) 
, the nature and properties of dark matter (e.g., \citealt{Viel_DM_2005PhRvD..71f3534V}, \citealt{Viel_DM_2013PhRvD..88d3502V}, \citealt{Irsic_FDM_2017PhRvL.119c1302I}, \citealt{Armengaud_DM_2017MNRAS.471.4606A}, 
\citealt{Rogers_DM_2021PhRvL.126g1302R})
and in combination with the cosmic microwave background (CMB, e.g. \citealt{Planck2020}) also inflation and neutrino masses  
(e.g., \citealt{Seljak_Cosmo_2006JCAP...10..014S}, \citealt{Palanque-Delabrouille_Neutrino_2015JCAP...11..011P}, \citealt{Yeche_Neutrino_2017JCAP...06..047Y}, \citealt{Palanque-Delabrouille_Neutrino_2020JCAP...04..038P}). %

The classical way of doing parameter inference with the {\lya} forest%
, as for any other cosmic tracer, relies on summary statistics of the underlying field, as they conveniently pick out a small number of relevant features from a much larger number of degrees of freedom of the full data. %
For the {\lya} forest, a number of summary statistics exists that have been accurately measured and effectively used for cosmological and astrophysical parameter inference. These include the line-of-sight (1D) transmission power spectrum (TPS hereinafter; %
e.g., \citealt{Croft_P1D_1998ApJ...495...44C}, \citealt{Chabanier_P1D_2019JCAP...07..017C}, \citealt{Walther_IGM_2019ApJ...872...13W}, \citealt{Boera_Inference_P1D_2019ApJ...872..101B}, \citealt{DESI_Lya_FFT-P1D_2023arXiv230606311R}, \citealt{DESI_Lya_QMLE-P1D_2023arXiv230606316K}), transmission PDF (TPDF hereinafter; e.g., \citealt{McDonald_FPDF_2000ApJ...543....1M}, \citealt{Bolton_FPDF_2008MNRAS.386.1131B}, \citealt{Viel_Inference_FPDF_2009MNRAS.399L..39V}, \citealt{Lee_FPDF_2015ApJ...799..196L}), wavelet statistics (e.g., \citealt{Meiskin_wavelets_2000MNRAS.314..566M}, \citealt{Theuns_wavelets_2000MNRAS.317..989T}, \citealt{Zaldarriaga_wavelets_2002ApJ...564..153Z}, 
\citealt{lidz2010},
\citealt{Wolfson_Inference_Wavelet_2021MNRAS.508.5493W}), curvature statistics (e.g., \citealt{Becker_Inference_curvature_2011MNRAS.410.1096B}, \citealt{Boera_Inference_curvature_2014MNRAS.441.1916B})%
, distributions of absorption line fits (e.g., %
\citealt{Schaye_bPDF_2000MNRAS.318..817S}, \citealt{Bolton_Inference_bPDF_2014MNRAS.438.2499B}, \citealt{Hiss_Inference_bPDF_2019ApJ...876...71H}, \citealt{Telikova_Inference_bPDF_2019ApJ...887..205T}, \citealt{Hu_Inference_LFI_2022MNRAS.515.2188H}) and combinations thereof (e.g., \citealt{Garzilli_Inference_FPDF_Wavelets_2012MNRAS.424.1723G}, \citealt{Gaikwad_IGM_2021MNRAS.506.4389G}).
While these provide accurate measurements of parameter values, they fail to capture the full information contained in the transmission field, thereby resulting in a loss of constraining power the full spectral data-sets have to offer.

Recently, deep learning approaches have become popular in the context of cosmological simulations and data analysis. Complex and resource-heavy conventional problems in cosmology have started to see fast, efficient, and demonstrably robust solutions in neural network (NN) based algorithms (see, e.g., \citealt{Review_AI_Cosmo_2023RPPh...86g6901M} for a recent review). %
Artificial intelligence has opened up a broad avenue for studies of the {\lya} forest as well. Cosmological analyses with the {\lya} forest generally demand expensive hydrodynamic simulations for an accurate modeling of the small-scale physics of the IGM. Deep learning offers alternative, light-weight solutions to such problems. For instance, \cite{Harrington_2022ApJ...929..160H} and \cite{Boonkongkird_LyAI-Net_2023arXiv230317939B} %
recently %
built %
U-Net based frameworks for directly predicting hydrodynamic quantities of the gas from computationally much less demanding, dark-matter-only simulations. 
A super-resolution generative model of {\lya}-relevant hydrodynamic quantities is presented in \citet{Jacobus_Superresolution_2023arXiv230802637J}, based on conditional generative adversarial networks (cGANs). These works greatly accelerate the generation of mock data for {\lya} forest analyses. Deep learning is also demonstrated to be a very effective methodology for a variety of tasks involving spectral, one-dimensional data-sets. %
\citet{Durovcikova_2020MNRAS.493.4256D} %
introduced a deep NN to reconstruct high-$z$ quasar spectra containing {\lya} damping wings. \citet{AutoencodingGalaxiesI_2022arXiv221107890M} and \citet{AutoencodingGalaxiesII_2023arXiv230202496L} describe a framework for generating, analysing, reconstructing and detecting outliers from SDSS galaxy spectra that consists of an autoencoder and a normalizing flow architecture. %
Recent works have shown immense potential of various deep NN methods for the analysis of the {\lya} forest. For example, a convolutional neural network (CNN) model to detect and characterize damped {\lya} systems (DLAs) in quasar spectra was introduced by \citet{Parks:2018}. Similarly, \citet{busca:2018} applied a deep CNN called ``QuasarNET'' for the identification (classification) and redshift-estimation of quasar spectra. \citet{Huang_2021MNRAS.506.5212H} constructed a deep learning framework to recover {\lya} optical depth from noisy and saturated {\lya} forest transmission. Later, \citet{Wang_2022MNRAS.515.1568W} applied the same idea to the reconstruction of the line-of-sight temperature of the IGM and detection of temperature discontinuities (e.g., hot bubbles). In %
neighbouring disciplines, deep learning is already identified as a reliable tool for %
field-level inference. For instance, a set of recent works (\citealt{Gupta_PhysRevD.97.103515}, \citealt{Fluri_PhysRevD.98.123518}, \citealt{Ribli_2019MNRAS.490.1843R}, \citealt{Kacprzak_PhysRevX.12.031029} among others) has established the superiority of deep learning techniques for cosmological inference directly from weak gravitational lensing maps over the classical two-point statistics of the cosmic density-field proxies. 

In this work we present {\lyanna} -- short for ``{\lya} Neural Network Analysis'' -- a deep learning framework for the analysis of the {\lya} forest. Here, we have implemented a 1D ResNet (a special type of CNN with skip-connections between different convolutional layers to learn the residual maps; \citealt{Resnet_2015arXiv151203385H}) called ``{\sansa}'' for inference of model parameters with {\lya} forest spectral data-sets, harvesting the full information carried by the transmission field. We perform non-linear regression on the thermal parameters of the IGM directly from the spectra containing the {\lya} forest absorption features that are extracted efficiently by our deep model. This architecture is trained in a supervised fashion using a large set of mock spectra from cosmological hydrodynamic simulations with known parameter labels to not only distinguish between two distinct parameter combinations but also pinpoint the exact location of a given spectral set in the parameter space. For better statistical reliability of our results, we employ a committee of 20 neural networks for the inference, combining the outputs via bootstrap aggregation \citep{Breiman_Bagging}. 
Finally, we build a likelihood model to perform inference on mock data-sets via Markov chain Monte Carlo (MCMC) and 
compare with classical summary statistics, namely a combination of TPS and TPDF,
showcasing the improvement we gain by working at field-level.

This paper is structured as follows. Section~\ref{sec:data} describes the simulations, the mock {\lya} forest spectra we use for training and testing our methodology, and the summary statistics we compare to. 
In Section~\ref{sec:neural-nets} we introduce the inference framework of {\sansa} with details of the architecture and its training. 
Our results of doing inference with {\sansa} and a comparison with the traditional summary statistics are presented and discussed in Section~\ref{sec:results}.
We conclude in Section~\ref{sec:conclusion} with a précis of our findings and an outlook.

\section{Simulations} \label{sec:data}
    In this section we introduce the hydrodynamic simulation used throughout this work as well as the post-processing approach we adopt to generate mock {\lya} forest spectra.
        \begin{figure*}
            \centering
            \includegraphics[width=\linewidth]{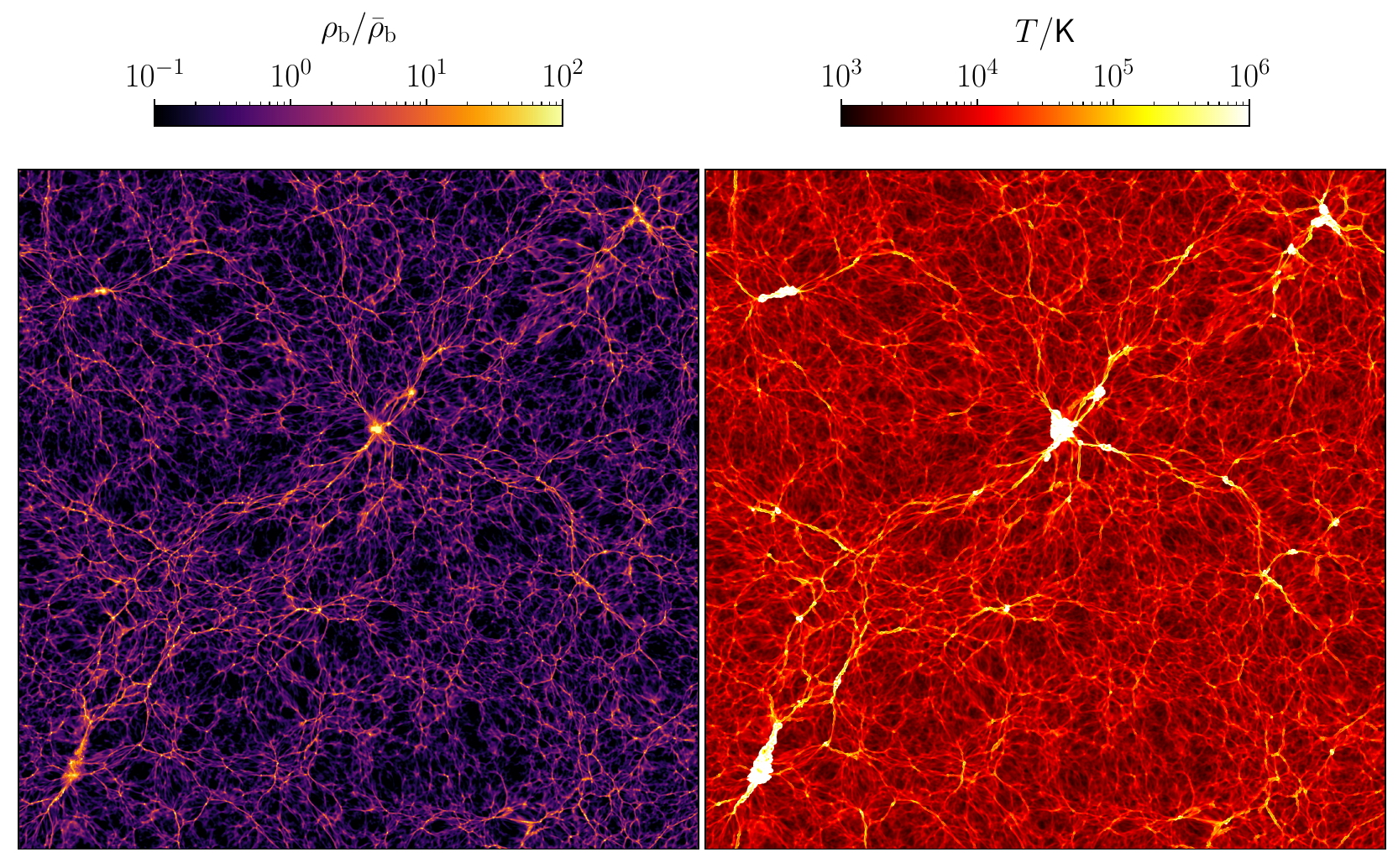}
            \caption{A slice (29 kpc thick) through our {\Nyx} simulation box (120 Mpc side-length, 4096 cells along a side) at $z = 2.2$ shown here for the gas overdensity (left) and temperature (right). A systematic relationship between both the hydrodynamic fields can be seen. %
            }
            \label{fig:slice-plot}
        \end{figure*}
    
    \subsection{Hydrodynamic Simulations} \label{sub:hydro-sims}
        We use a {\Nyx} cosmological hydrodynamical simulation snapshot %
        generated for {\lya} forest analyses (see \citealt{Walther_Nyx_2021JCAP...04..059W}) %
        to create the mock data used for various purposes in this work. {\Nyx} is a relatively novel hydrodynamics code based on the \textsc{AMReX} framework and simulates
        an ideal gas on an Eulerian mesh interacting with dark matter modeled as Lagrangian particles. While adaptive mesh refinement (AMR) is possible and would allow better treatment of overdense regions, we used a uniform grid here as the {\lya} forest only traces mildly overdense gas, rendering AMR techniques inefficient. Gas evolution is followed using a second-order accurate scheme (see \citealt{Almgren_Nyx_2013ApJ...765...39A} and \citealt{Lukic_2015MNRAS.446.3697L} for more details).
        In addition to solving the Euler equations and gravity, {\Nyx} also models the main physical processes required for an accurate model of the {\lya} forest. The chemistry of the gas is modeled following a primordial composition of H and He. Inverse Compton cooling of the CMB is taken into account as well as the updated recombination, collisional ionization, dielectric recombination and cooling rates from \cite{Lukic_2015MNRAS.446.3697L}. All cells are assumed to be optically thin to ionizing radiation and a spatially uniform ultraviolet background (UVB) is applied according to the late reionization model of \cite{Onorbe:2018}, where heating rates have been modified by a fixed factor $A_\text{UVB}$ affecting the thermal history and thus pressure smoothing of the gas. Here, we use a simulation box at $z=2.2$ with 120 Mpc side-length and $4096^3$ volumetric cells (``voxels'') and dark matter particles, motivated by recent convergence analyses (\citealt{Walther_Nyx_2021JCAP...04..059W} and \citealt{Chabanier:2023}). The cosmological parameters of the box are $h = 0.7035$, $\omega_{\text{m}} = \Omega_{\text{m}}h^2 = 0.1589$, $\omega_{\text{b}} = \Omega_{\text{b}}h^2 = 0.0223$, $A_{\mathrm{s}} = 1.4258$, $n_{\mathrm{s}}=1.0327$, $A_{\mathrm{UVB}}=0.9036$. 

        \begin{figure*}
            \centering
            \includegraphics[width = \linewidth]{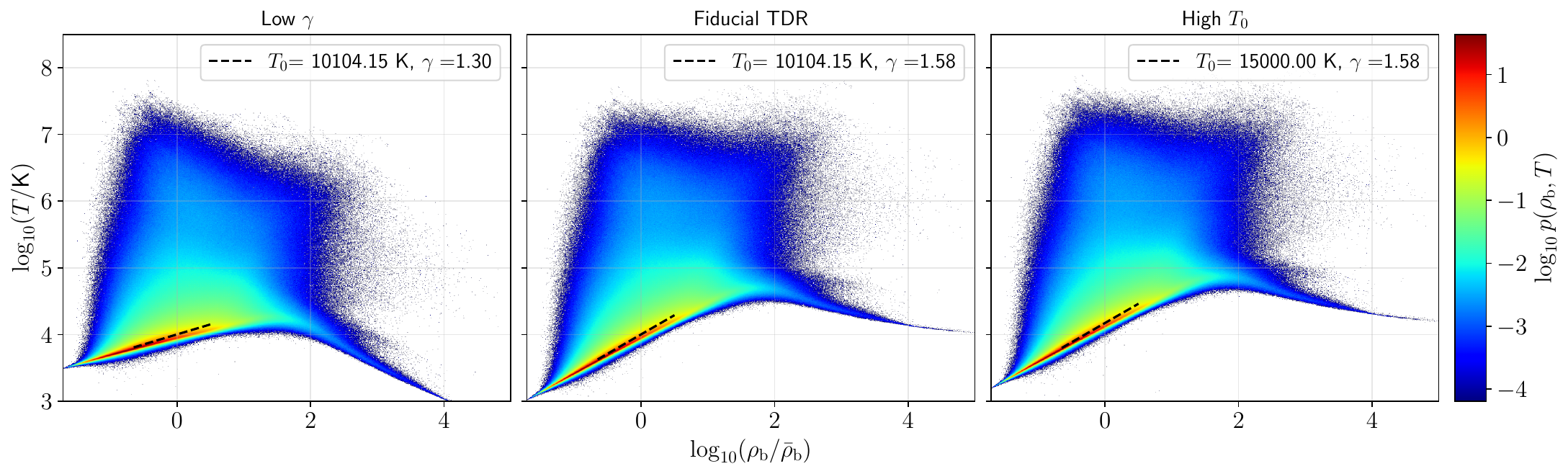}
            \caption{Joint, volume-weighted distribution of the temperature and density of baryons in our simulation at $z = 2.2$. \textbf{Center}: the fiducial $(T_0,\gamma)$ parameter case (values obtained by fitting a power-law through gas temperatures and densities). \textbf{Left}: the case with rescaled temperatures for a lower $\gamma$ than the fiducial and the same $T_0$. This can be seen to affect the slope of the TDR by a twisting of the 2D distribution. \textbf{Right}: the case with rescaled temperatures for a higher $T_0$ than the fiducial and the same $\gamma$. The entire distribution is shifted along the $\log_{10}(T/$K) axis, keeping the shape the same. (Note that the color-bar label is using shorthand for $p(\log_{10}(\rho_\mathrm{b}/\bar{\rho}_\mathrm{b}), \log_{10}(T/$K$))$.)
            }
            \label{fig:tdr_hist}
        \end{figure*}

        During the epoch of reionization the ionizing UV radiation from star-forming galaxies heats up the intergalactic gas as well. Afterward, as the universe expands, the IGM cools down mostly adiabatically with subdominant non-adiabatic contribution, e.g., from inverse Compton scattering off of the CMB and recombination of the ionized medium, as well as heating due to photoionization and gravitational collapse. The bulk of this gas is diffuse (relatively cool with $T<10^5$ K and mildly overdense with $\log_{10} (\rho_\mathrm{b}/\bar{\rho}_\mathrm{b})<2$, 
        contained mostly in cosmic voids, sheets, and filaments; see, e.g., \citealt{Matizzi_CosmicWeb_2019MNRAS.486.3766M}) and imprints the {\lya} forest absorption features on to QSO spectra. The IGM at $z\sim 2.2$ can be further classified into subdominant phases such as warm-hot intergalactic medium (WHIM; $T>10^5$ K and $\log_{10} (\rho_\mathrm{b}/\bar{\rho}_\mathrm{b})<2$), condensed halo ($T<10^5$ K and $\log_{10} (\rho_\mathrm{b}/\bar{\rho}_\mathrm{b})>2$) and warm halo or circumgalactic medium (WCGM; $T>10^5$ K and $\log_{10} (\rho_\mathrm{b}/\bar{\rho}_\mathrm{b})>2$). At this redshift, the effects due to the inhomogeneous UVB of the reionization of H and He are considered to be 
        small (e.g., \citealt{Onorbe_2019MNRAS.486.4075O, UptonSanderbeck_2020MNRAS.496.4372U}) and we ignore them for this work. The diffuse IGM component in our cosmological simulation exhibits a tight power-law relation in temperature and density \citep{Hui_Gnedin_1997MNRAS.292...27H, McQuinn_powerlaw_2016MNRAS.456...47M} that is classically characterized by
        
        \begin{equation}\label{eqn:tdr}
            T = T_0 \bigg( \frac{\rho_\text{b}}{\bar{\rho}_\text{b}} \bigg) ^{\gamma - 1},
        \end{equation}
        where $\bar{\rho}_\text{b}$ is the mean density of the gas, and %
        $T_0$ (a temperature at the mean gas density) and %
        $\gamma$ (adiabatic power-law index) are the two free parameters of the model. Indeed, a strong systematic $\rho_\mathrm{b}$-$T$ relationship is visually apparent in a slice through our simulation box (\autoref{fig:slice-plot}). We perform a linear least-squares fit of the above relation through our simulation in the range $-0.5 < \log_{10}(\rho_\text{b} / \bar{\rho}_\text{b}) < 0.5$ and $\log_{10}(T/\text{K}) < 4$. The best-fit (fiducial) values are $T_0 = 10104.15$ K %
        and $\gamma = 1.58$. While a range of works  have demonstrated the potential of using different summary statistics of the {\lya} forest as probes to measure $T_0$ and $\gamma$ \citep[see e.g.][]{Gaikwad_IGM_2020MNRAS.494.5091G}, in this work we highlight a first field-level framework for inference of these two thermal parameters of the IGM.
        
        The following strategy is adopted for sampling the parameter space of $(T_0, \gamma)$ to produce labeled data for the supervised training of the inference machine. Both the parameters are varied by a small amount at a time, $\log T_0 \to \log T_0 + \log x$ and $\gamma \to \gamma + y$ to obtain a new TDR.
        We then rescale %
        the simulated temperatures at every cell of the simulation by $T \to x\cdot(\rho_\text{b}/\bar{\rho}_\text{b})^y \cdot T$ at fixed densities $\rho_\text{b}$ to appropriately incorporate the 
        scatter off the TDR into our mock data, effectively conserving the underlying $T$-$\rho_{\mathrm{b}}$ distribution rather than drawing from a pure power-law. This procedure is illustrated in \autoref{fig:tdr_hist} with the help of the full 2D histograms of temperature and density for two individual parameter rescalings as well as the fiducial case. %

    \subsection{Mock Lyman-$\alpha$ Forest} \label{sub:mocks}

        \begin{figure*}
            \centering
            \includegraphics[width = \linewidth]{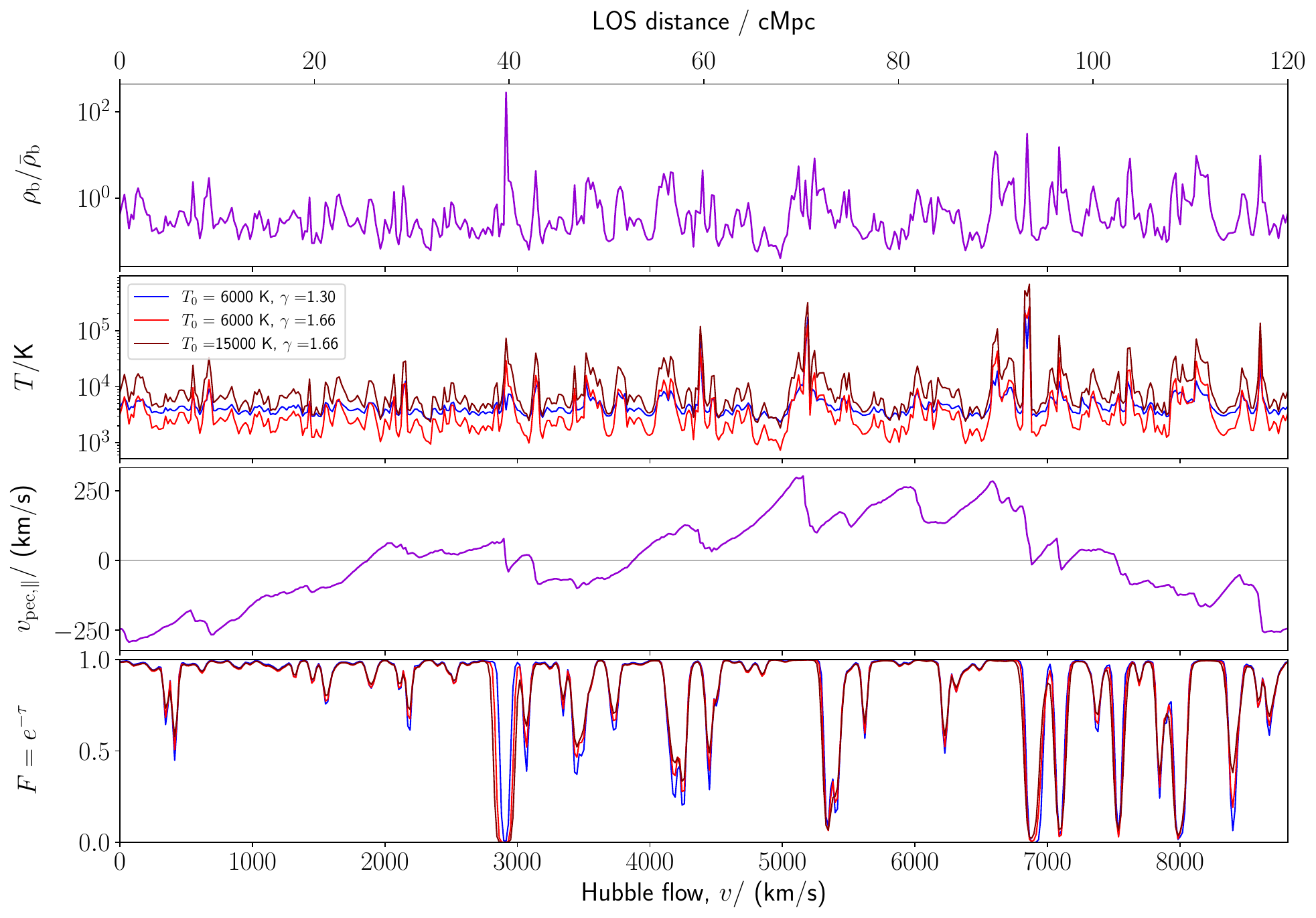}
            \caption{Simulated baryon overdensities, 
            temperatures, line-of-sight peculiar velocity, and transmission along an example skewer through our box for three different temperature-density relations -- one pair each for fixed $T_0$ and fixed $\gamma$ -- to indicate characteristic variations \emph{w.r.t} the two TDR parameters.%
            The line shifts due to the peculiar velocity component ($v_{\text{pec},\parallel}$) can be easily noticed. Since the rescaling of the temperatures depends on the densities, inhomogeneous differences are seen in the skewer temperatures in the cases with varying $\gamma$. The absorption lines are broader where the temperatures are higher and the amount of broadening is also a function of densities being probed through $\gamma$.}
            \label{fig:example-skewer}
        \end{figure*}
        In order to simulate the {\lya} forest transmission $F = e^{-\tau}$ ($\tau$ being the optical depth), %
        we first choose $10^5$ random lines of sight (LOS, a.k.a. skewers) parallel to one of the Cartesian sides (e.g. %
        Z-axis) of the box by picking all consecutive 4096 voxels along that axis while keeping the other two coordinates (X and Y) %
        fixed at a time. 
        The {\lya} optical depth at an output pixel in a spectrum is calculated %
        from the information of the density, temperature, and the LOS component of gas peculiar velocity ($v_{\mathrm{pec},\parallel}$) at each corresponding voxel along the given skewer.  %
        Here, the gas is reasonably assumed to be in ionization equilibrium among the different species of H and He and further that He is almost completely (doubly) ionized at $z \sim 2.2$ (i.e. $x_{\HeIII} \approx 1$; \citealt{Miralda-Escude_HeII_Reion_2000ApJ...530....1M}, \citealt{Becker_Reion_2015MNRAS.447.3402B}) in order to estimate the neutral H density, $n_{\HI}$ for each of those voxels. 
        The {\lya} optical depth at a pixel with Hubble velocity $v$ and gas peculiar velocity $v_{\mathrm{pec},\parallel}$ is estimated as 
        \begin{equation}
            \tau (v) = \frac{\pi e^2 \lambda_{\mathrm{Ly}\alpha} f_{lu}}{m_\mathrm{e} c H(z)} \int n_{\HI}(v') \phi_\mathrm{D} \big( v'; v+v_{\mathrm{pec},\parallel}, b \big) dv',
        \end{equation}
        where the rest-frame $\lambda_{\mathrm{Ly}\alpha} = 1215.67$ \AA, the {\lya} oscillator strength $f_{lu} = 0.416$, and
        \begin{equation}
            \phi_\mathrm{D}(v; v_0, b) \equiv \frac{1}{b\sqrt{\pi}} \exp \Bigg[ - \bigg( \frac{v - v_0}{b} \bigg)^2 \Bigg]
        \end{equation}
        is the Doppler line profile with the temperature-dependent broadening parameter $b = \sqrt{2k_\mathrm{B}T/m_\mathrm{H}}$. These $\tau$ values are additionally rescaled by a constant factor such that the mean {\lya} transmission in our full set of $10^5$ skewers matches its observed value of $\bar{F}_\mathrm{obs} = 0.86$, compatible with \citet{Becker_2013MNRAS.430.2067B}. Four our simulation, we performed tests with a simplified, approximate model of lightcone evolution along the LOS (Appendix~\ref{app:lightcone}) and found negligible impact on the performance of our inference framework. Therefore, we ignore any small lightcone evolution along our skewers and use the snapshot of the simulation for creating mock {\lya} forest spectra, assuming a constant TDR for simplicity.
        
        To mimic observational limitations and minimize the impact of numerical noise on small scales in the simulations, we restrict the Fourier modes within the spectra to $k < k^*$, $k^* = 0.18$ s km$^{-1}$.
        This is effectively achieved by smoothing them with a spectral resolution kernel of $R_\mathrm{FWHM} \approx 11,000$ and additionally rebinning them by 8-pixel averages, matching the Nyquist sampling limit.
        The final size of a spectrum in our analysis is thus 512 pixels. %

        \begin{figure}
            \centering
            \includegraphics[width=1\linewidth]{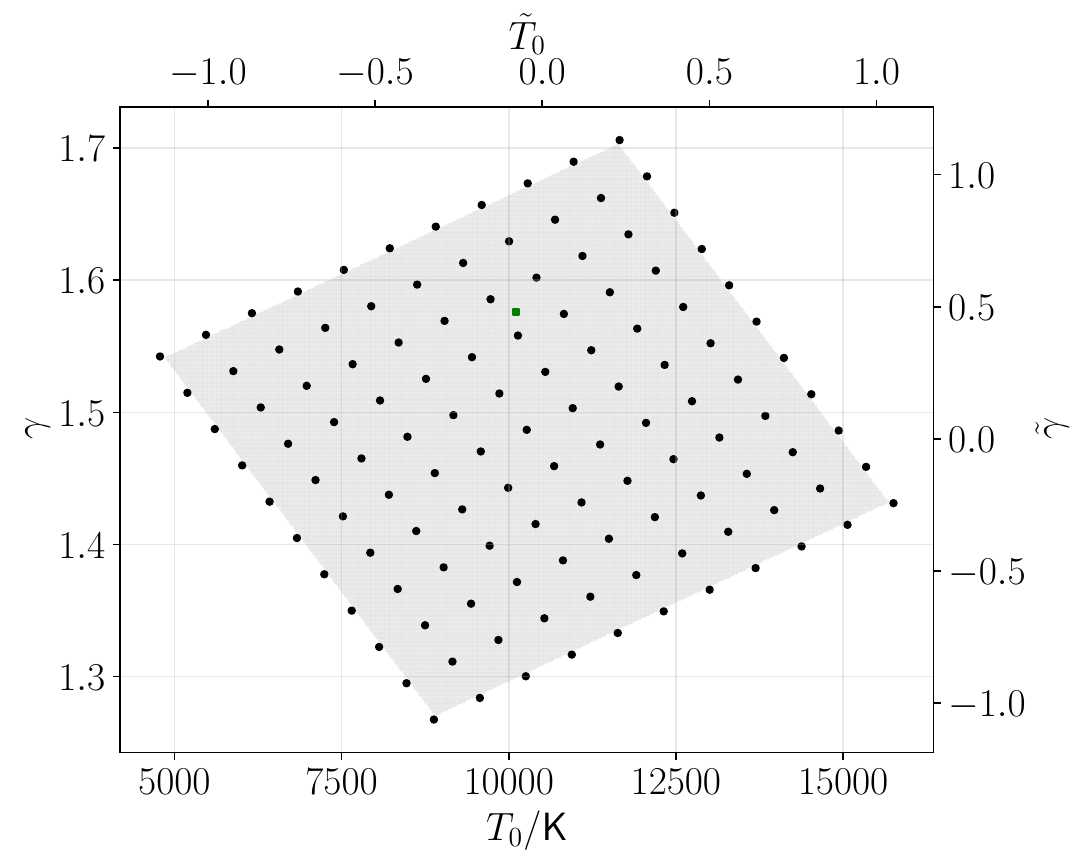}
            \caption{Our sample of thermal models for the various training and test purposes that contains 11$\times$11 (121) distinct $(T_0,\gamma)$ combinations. The fiducial TDR parameter combination is depicted as the green square. The rescaled $(\tilde{T}_0, \tilde{\gamma})$ axes are also shown for context (see Section~\ref{sub:lyanna-overview}). 
            This is a uniform grid in a parametrization that captures the well known degeneracy direction in the $(T_0,\gamma)$ space. The grey-shaded area shows our prior range in this parameter space. Please refer to Appendix~\ref{app:orthogrid} for more details regarding the strategy used for this sampling.}
            \label{fig:orthogrid_space}
        \end{figure}
         When sampling the $(T_0, \gamma)$ space, new mock spectra are produced for each parameter combination with the new (rescaled) temperatures and the original densities and line-of-sight peculiar velocities along the same set of skewers.

         An example skewer is shown in \autoref{fig:example-skewer} for three different TDRs. Since the temperature rescaling is a function of $\rho_\text{b}/\bar{\rho}_\text{b}$ through $\gamma$, characteristic differences in skewer temperatures and {\lya} transmission between cases with varying $\gamma$ are visible. Changes in $T_0$ result in a homogeneous broadening of the absorption lines, whereas changes in $\gamma$ (for $\gamma>1$)\footnote{$\gamma = 1$ would mean the diffuse IGM is roughly isothermal and $\gamma < 1$ would lead to an ``inverted TDR,'' where underdensities are hotter than overdensities, as proposed e.g. by \cite{Bolton_FPDF_2008MNRAS.386.1131B}.
         }, 
         depending on the underlying overdensity being probed, result in larger or smaller broadening (i.e., generally, the shallower lines are broadened less than the deeper ones). We expect our convolutional architecture to be able to pick up such features in order to discriminate between thermal models.
         In this work, we sample a grid of 11$\times11$ $(T_0,\gamma$) combinations as shown %
         in \autoref{fig:orthogrid_space} -- for each of which we have the same set of $10^5$ physical skewers %
         -- for %
         training and testing our deep learning machinery. This grid is oriented in a coordinate system that captures the well-known degeneracy direction in the $(T_0,\gamma)$ space as identified in many TPS analyses (e.g. \citealt{Walther_IGM_2019ApJ...872...13W}) and is motivated by the heuristic argument that it is easiest to train a neural network for inference with an underlying parametrization that captures the most characteristic variations in the data. The exact sampling strategy is further described in Appendix~\ref{app:orthogrid}. We use the grey-shaded region in \autoref{fig:orthogrid_space} as our prior range of parameters having a uniform prior distribution in all our further analyses. %
         \begin{figure*}
            \centering
            \includegraphics[width=\linewidth]{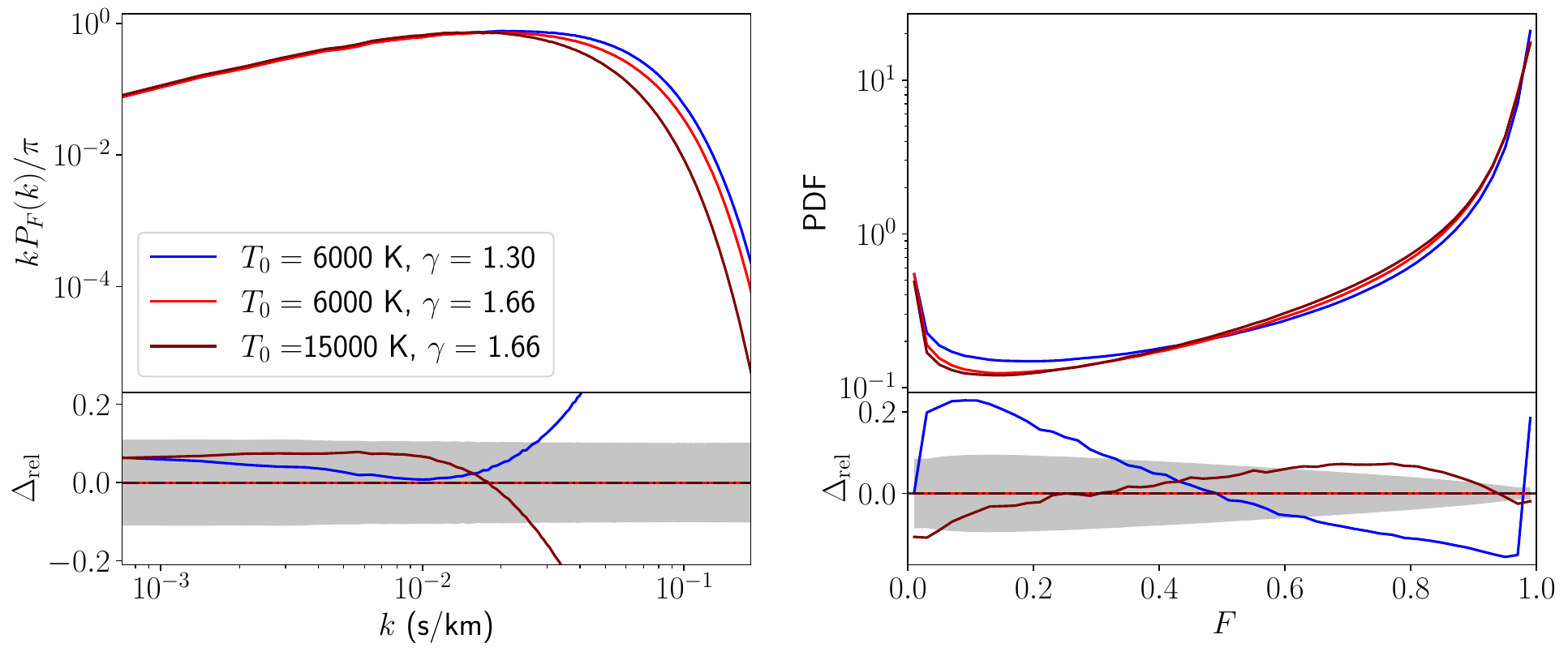}
            \caption{Transmission power spectrum (left) and transmission PDF (right) computed from our set of $10^5$ skewers for three TDR parameter combinations, the same as in \autoref{fig:example-skewer}. The fractional differences between different TDR cases are shown in the corresponding bottom panels with the grey-shaded areas as $1\sigma$ uncertainty ranges (bands drawn from discrete $k$-modes in TPS and discrete histogram bins in TPDF), equivalent to 100 spectra. 
            }\label{fig:example_summaries}
        \end{figure*}

    \subsection{Summary Statistics}\label{sub:summaries}
        \begin{figure}
            \centering
            \includegraphics[width=1\linewidth]{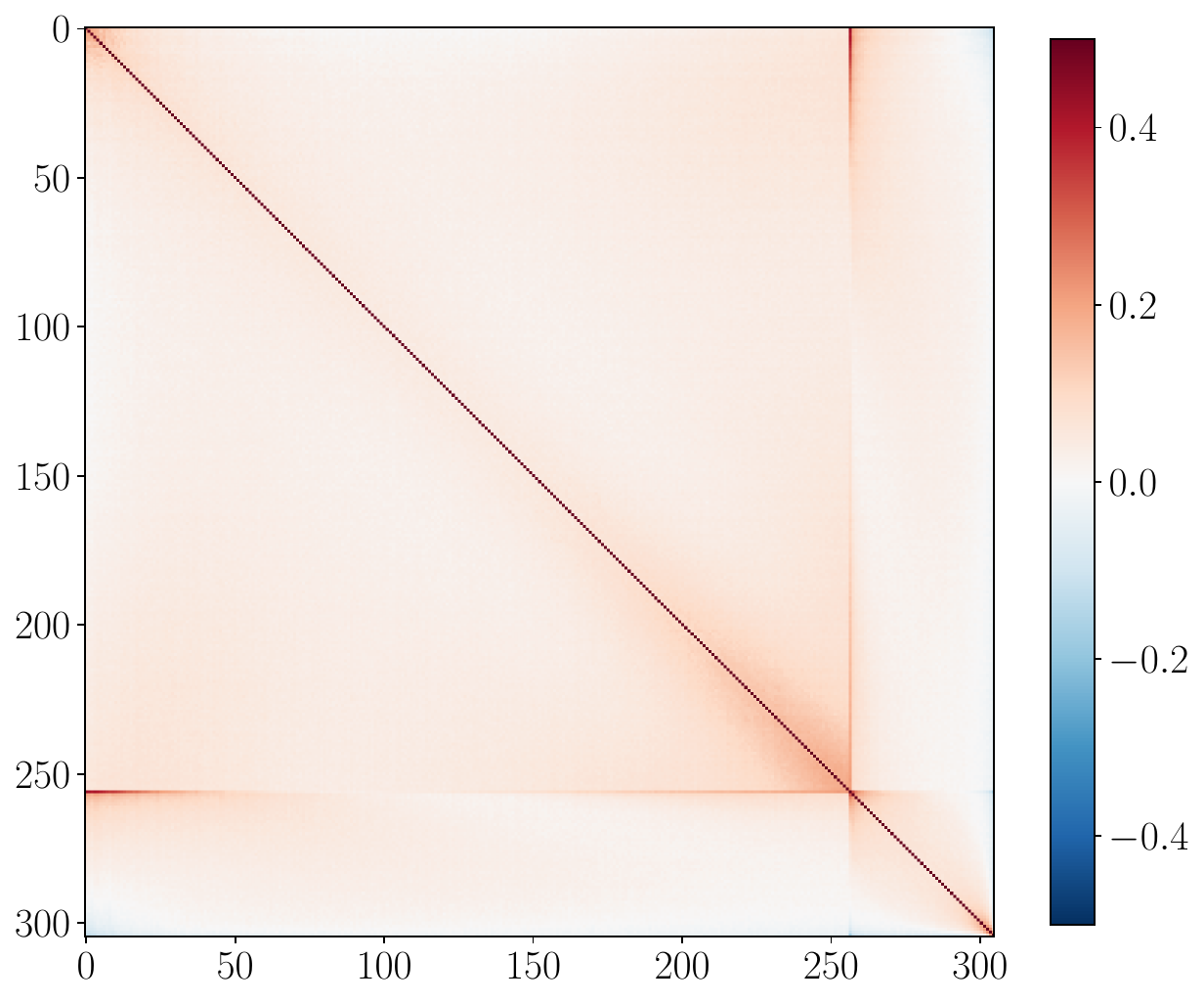}
            \caption{Correlation matrix of the joint summary vector (the first 256 entries being TPS and the later 49 being TPDF) estimated from our set of $10^5$ mock spectra for the fiducial thermal model. Notice the mild correlations within the individual summaries and the cross-correlations between the two summaries.}
            \label{fig:corr-matrix-summary}
        \end{figure}
        We consider two summary statistics -- TPS and TPDF -- of the {\lya} forest in this work for demonstrating the benefit of field-level inference. %
        The TPS is defined here as the variance of the transmitted ``flux contrast'' $\delta_F \equiv (F -\bar{F}) / \bar{F}$ in Fourier space, i.e., $P_F(k) \sim \langle \tilde{\delta}(k)^* \cdot \tilde{\delta}(k) \rangle$. %
        For a consistent comparison of inference outcomes, we apply the same restriction $k<0.18$ s km$^{-1}$ as in the input to our deep learning machinery (see Section~\ref{sub:mocks}). %
        To obtain the TPDF, we consider the histogram of the transmitted flux $F$ in the full set of skewers over 50 bins of equal width between 0 and 1. For the likelihood analysis with the TPDF we leave the last bin out as it is fully degenerate with the rest due to the normalization of the PDF. The mean TPS and TPDF computed from the $10^5$ skewers for three different TDR parameter combinations are shown in \autoref{fig:example_summaries} along with the relative differences in both the statistics between pairs of TDR models. The uncertainty range shown as a grey band therein corresponds to a $1\sigma$ equivalent of 100 spectra. The TPS follows a power-law increase for small $k$ and exhibits a suppression of power at larger $k$ (smaller physical scales) due to deficiency of structures as well as the thermal broadening of lines. 
        The variations in the thermal parameters that effectively result in the broadening of absorption lines amount to a shift in this turnover scale toward smaller $k$-modes\footnote{Recall that the mean transmission $\bar{F}$ is kept fixed for our spectra; for fixed UVB -- hence varying $\bar{F}$ -- the quantitative variations in the power spectrum \emph{w.r.t.} the thermal parameters would be different, especially for small $k$.}. Equivalently, the broader lines have shallower depths (in this low-density regime of the curve of growth), which in turn results in a transfer of probability from smaller ($F\lesssim0.4$) to larger ($F\gtrsim0.6$) transmission. We also compute the joint covariance matrix of the concatenated summary vector of TPS and TPDF from our full set of mock spectra by the estimator
        \begin{equation}
            \mathbf{C_s} = \frac{1}{N-1} \sum_{i=1}^{N} (\mathbf{s}_i - \mathbf{\bar{s}})^{T} (\mathbf{s}_i - \mathbf{\bar{s}}),
        \end{equation}
        where $N=10^5$ and $\mathbf{s}$ is a vector of which the first 256 entries are the Fourier modes in the TPS ($k < k^*$) and the later 49 entries are the bins in the TPDF, $F\in [0,1)$.  The joint correlation matrix for a thermal model from our sample is shown in \autoref{fig:corr-matrix-summary}. Mild correlations among relatively close entries within the TPS or the TPDF can be observed as well as moderate cross-correlations between the two summary statistics.
        
        For inference with these summary statistics, we cubic-spline interpolate both (the TPDF per histogram bin and the TPS per discrete $k$-mode) as a function of the parameters $(T_0, \gamma)$ to obtain an emulator over our prior range as depicted in \autoref{fig:orthogrid_space} where we assume a flat (uniform) prior in both the parameters. We verified that choosing a different interpolation scheme, e.g. linear, does not strongly affect the results of the inference.

\section{Field-level Inference Machinery} \label{sec:neural-nets}

    As described in Section~\ref{sec:data}, we have simulated {\lya} forest absorption profiles (spectra) from hydrodynamic simulations having known thermal $(T_0,\gamma)$ parameter values. The aim of our machinery is to learn the characteristic variations in the spectra (i.e., at field-level) \emph{w.r.t.} those parameters in order first to distinguish between two adjacent thermal models and ultimately also to provide an uncertainty estimate as well as a point estimate of the parameter values whereby Bayesian inference can be performed. Thus framed, this is a very well-suited problem for application of supervised machine learning. The output of a fully-trained deep neural network can be used as a model (emulator) for a newly-learned, optimal ``summary statistic'' of the {\lya} transmission field that is fully degenerate with the thermal parameters\footnote{Meaning that the machine can summarize the field most optimally (informed by the full data) into $N_\mathrm{par}$ values that can be directly mapped to the actual parameters of interest.}, hence carrying most of the relevant information about them that the full field offers. 

    In the following we describe our framework in detail, with special focus on the neural network architecture and training.

    \subsection{Overview} \label{sub:lyanna-overview}
        The general structure of inference with {\lyanna} entails a feed-forward 1D ResNet neural network called ``{\sansa}'' that connects an underlying input information vector (transmission field) to an output ``summary vector'' that can be conveniently mapped to the thermal parameters $(T_0,\gamma)$. Ideally, we expect this summary vector to be a direct actual estimate of the parameters itself, however, due to a limited prior range of thermal models available for training, a systematic (quantifiable) bias is observed in the pure network estimates (see Appendix~\ref{app:biases}). Nonetheless, these estimates can be mapped to the parameters via a tractable linear transformation. For brevity our network encompasses this mapping\footnote{Although this linear map is part of our network, it is not fitted during back-propagation in order to avoid the bias due to the prior limits.} such that its output is a direct estimate of the parameters $\mathbf{\tilde{\pi}} = (\tilde{T}_0$, $\tilde{\gamma})$. As an estimate of its own uncertainty of a given prediction, the network also returns a parameter covariance matrix $\tilde{\mathbf{C}}$. Since our two parameters have dynamic ranges different by orders of magnitude, we linearly rescale them as %
        $T_0 \to \tilde{T}_0$ and $\gamma \to \tilde{\gamma}$, to fall in the same range $\sim(-1, 1)$. %
        This bijective %
        mapping ensures numerical stability of the point estimates by the networks and is a common practice for deep learning regression schemes. The output covariance matrix (and its inverse) must be positive-definite as a mathematical requirement. This is ensured in our framework in a way %
        similar to \cite{Fluri2019_PhysRevD.100.063514} by a Cholesky decomposition of the form 
        \begin{equation}\label{eqn:cholesky}
            \tilde{\mathbf{C}}\mathbf{^{-1}} = \mathbf{L} {\mathbf{L}}^T,
        \end{equation}
        where $\mathbf{L}$ is a lower triangular matrix. Our network predicts the three independent components of that matrix ($\log {L_{11}}$, $\log {L_{22}}$, $L_{12}$ to further ensure uniqueness\footnote{Since $\tilde{C}^{-1}_{ii} = (\pm L_{ii})^2$, we explicitly choose the positive branch of the Cholesky coefficients (a unique, one-to-one mapping) via the lognormal transformation, for numerical stability reasons.})
        rather than the covariance matrix directly. The network is optimized following a Gaussian negative log-likelihood loss (hereinafter NL3), %
        \begin{equation}\label{eqn:nl3}
            \mathcal{L}(\tilde{\pi}) = \log \vert \tilde{\mathbf{C}} \vert + (\tilde{\pi} - \hat{\pi}) \tilde{\mathbf{C}}\mathbf{^{-1}} (\tilde{\pi} - \hat{\pi})^T,
        \end{equation}
        where $\hat{\pi} = (\hat{T}_0, \hat{\gamma})$ are the true parameter labels. 
        This can be seen as an extension of the conventional mean-squared error, MSE $=(\tilde{\pi} - \hat{\pi})(\tilde{\pi} - \hat{\pi})^T$, in the presence of a network-estimated covariance. It can be noted that the covariance matrix does not have any labels and is primarily a way to regularize network predictions under a Gaussian likelihood assumption. %

    \subsection{Architecture}\label{sub:architecture}
        \begin{figure*}
            \centering
            \includegraphics[width = \linewidth]{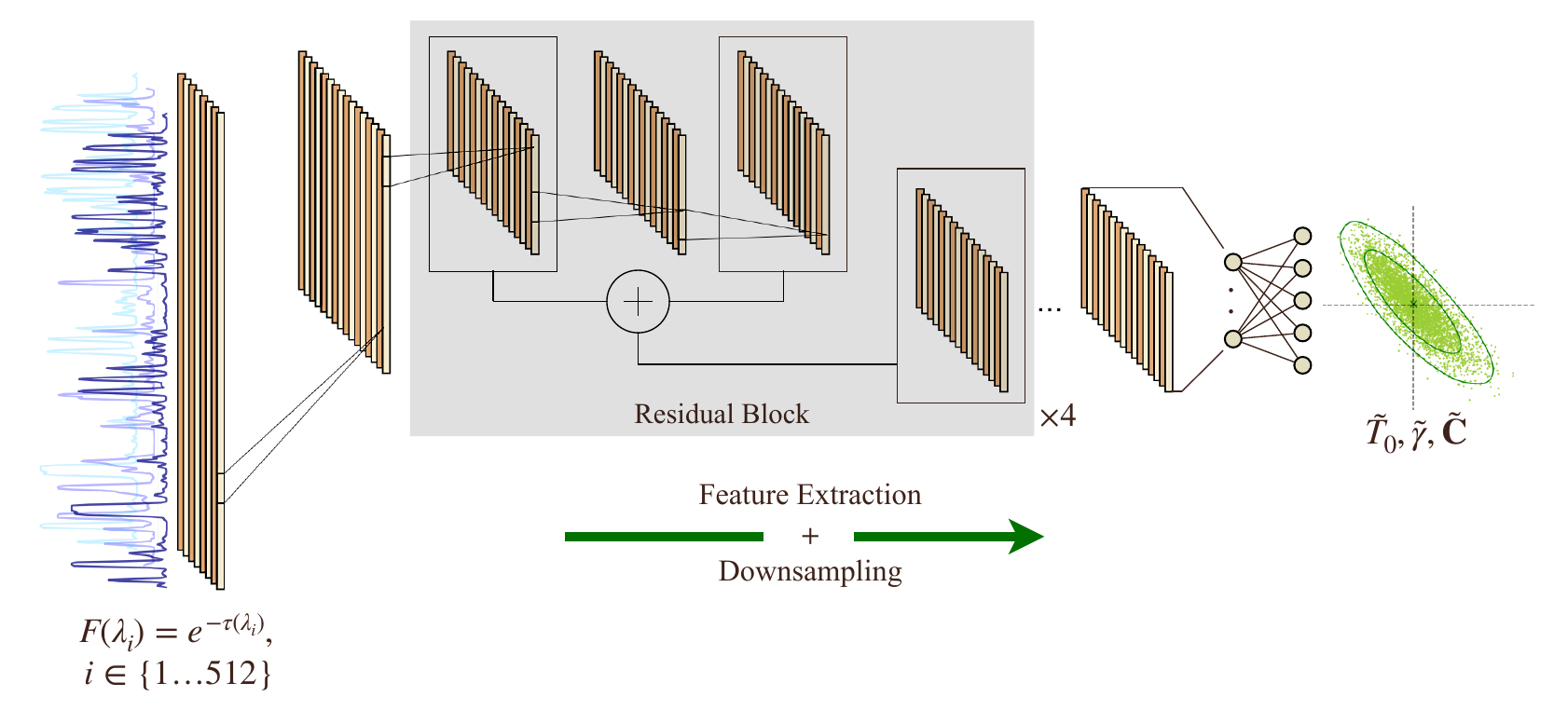}
            \caption{A schematic representation of the architecture of {\sansa}. It comprises a 1D residual convolutional neural network with four residual blocks in series for extracting crucial features from the spectra of size 512 pixels, followed by a fully-connected %
            layer to map the outcome to the parameter point predictions $\mathbf{\tilde{\pi}}$ and a covariance estimate $\mathbf{\tilde{C}}$ over 5 output neurons.}
            \label{fig:sansa-schematic}
        \end{figure*}
\begin{figure}
            \centering
            \includegraphics[width = 0.45\linewidth]{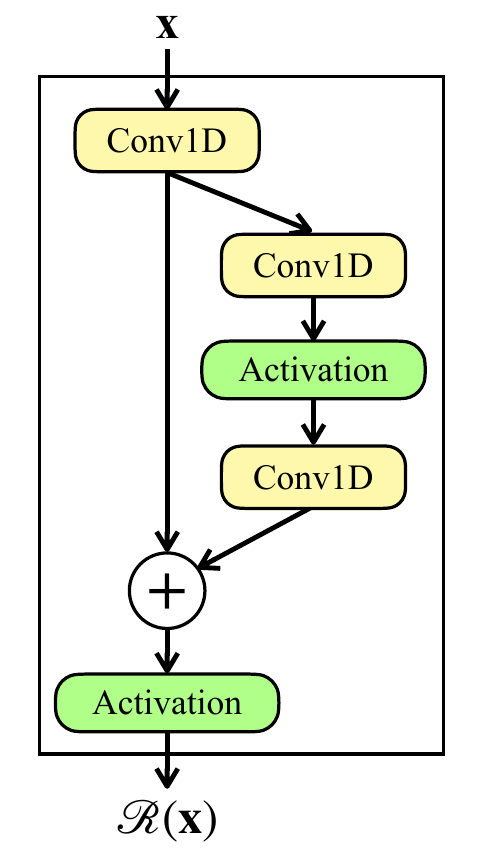}

            \caption{A residual block in {\sansa}. An input vector $\mathbf{x}$ is first passed through a convolutional layer and a copy of the output tensor is made which consecutively goes through a pair of convolutional layers introducing nonlinearity, 
            all the while preserving the shape of the output tensor. The outcome is then algebraically added to the earlier copy (i.e. a parallel, identity function) and the sum is passed through a nonlinear activation to obtain the final outcome of the block. The latter two convolutional layers thus learn a residual nonlinear mapping. 
            (Note that a zero-padding is applied during all convolutions in order to preserve the feature shape in the subsequent layers through the network.)
            }
            \label{fig:res-block}
        \end{figure}

        We build our architecture using the open-source \textsc{Python} package \textsc{TensorFlow/Keras} \citep{chollet2015keras}.
        The neural network for field-level inference with {\lyanna} is called {\sansa} and consists of a 1D ResNet \citep{Resnet_2015arXiv151203385H},
        a residual convolutional neural network that extracts useful features from spectra and turns them into a ``summary'' vector which can then be used for inference of model parameters. 
        \autoref{fig:sansa-schematic} shows a schematic representation of the architecture of {\sansa}. The input layer consists of 512 spectral units. %
        This input is passed through four residual blocks in series with varying numbers of input/output units, each block having the same computational structure as illustrated in \autoref{fig:res-block}. Each residual block is followed by a batch-normalization and an average-pooling layer to downsample the data vectors consecutively. A ResNet architecture is particularly attractive because of its ease of convergence during training owing to predominantly learning the residual mappings that could conveniently be driven to zero if identity maps are the most optimal in intermediate layers. This is achieved by introducing ``skip-connections'' in a sequential convolutional architecture that fulfil the role of identity (linear) functions. The residual blocks can more easily adapt to those linear mappings than having to train non-linear layers to mimic them. A special advantage of the skip-connections is that they do not introduce more parameters than a sequential counterpart. Our neural network has a total of 136,784 trainable parameters that are tuned via back-propagation. We use the \textsc{TensorFlow} in-built leaky ReLU (rectified linear unit) function for all the non-linear activations in the residual blocks with the negative-slope of 0.3. The resultant set of feature-tensors is flattened into a single vector of size 128 and mapped to the output vector $(\tilde{T}_0, \tilde{\gamma}, \log L_{11}, \log L_{22}, L_{12})$ with a fully-connected, unbiased  %
        linear layer. We regularize the network kernels with a very small L2 weight decay ($\mathcal{O}(10^{-8})$ in the convolutional layers, $\mathcal{O}(10^{-7})$ in the fully connected layer). We also use a dropout \citep{Dropout_Srivastava_2014}  of 0.09 after each residual block during training for encouraging generalization\footnote{Note that this value of the dropout rate $p$ is consistent with the \textsc{Keras} convention, i.e., the fraction of input layer units to drop, unlike the original definition by \citealt{Dropout_Srivastava_2014} where $p$ is the probability of the output of a given layer unit being propagated further in the network.}. 

    \subsection{Training}\label{sub:training}
        All convolutional kernels in {\sansa} are initialized following the approach of \citet{Glorot_Normal} and the weights in the final linear layers are initialized similarly to \citet{He_Normal}. After a preliminary convergence test \emph{w.r.t.} training data-set size, we chose a training set consisting of 10,000 distinct spectra from each thermal model in our sample. We also have a separate validation set for monitoring overfitting \emph{viz.} 2/5 the size of the training set with an equivalent distribution of spectra among thermal models. 
        The network is trained by minimizing the NL3 loss function in Eq.~\eqref{eqn:nl3}. Additionally, three other metrics are monitored during the training: $\log \vert \tilde{\mathbf{C}} \vert$, $\chi^2 = (\tilde{\pi} - \hat{\pi}) \tilde{\mathbf{C}}\mathbf{^{-1}} (\tilde{\pi} - \hat{\pi})^T$, and the MSE. Notice that the loss function is simply the sum of the first two metrics. 

        The training is performed by repeatedly cycling through the designated training data-set in randomly chosen batches of a fixed size. Each cycle through the data is deemed an ``epoch'', and each back-propagation action on a batch is termed a ``step of training''. Since the spectra follow periodic boundary conditions, a cyclic permutation of pixels (``rolling'') is mathematically allowed and leads to no alteration of underlying physical characteristics (e.g., thermal parameters $T_0, \gamma$). This is also true for reverting the order of the pixels (``flipping''). These are some of the modifications that augment the existing training set and we expect our network to be robust against. 
        Therefore, at every epoch we apply a uniformly randomly sampled amount (in number of pixels) of rolling and a flipping with 50\% probability to each of the training spectra, on-the-fly. (Note that the validation set is not augmented on-the-fly because we would like to compare the generalization of the network predictions at different epochs for the same set of input spectra.)
        
        We expect the $\chi^2$ metric to optimally take the value $\sim N_\mathrm{par}$ %
        because of the underlying Gaussian assumption (in our case $N_\mathrm{par} = 2$). The improvement of the network during training is then in large parts due to a decrement in $\log \vert \tilde{\mathbf{C}} \vert$ which indicates that the network becomes less uncertain of its estimates as the training progresses. The state of a network is said to be improving if the value of NL3 decreases and the network $\chi^2$ remains close to 2  for data unseen during back-propagation, the validation set. Therefore, we deem the best state of the network to be occuring at the epoch during training at which the validation NL3 is minimal while the validation $\vert \chi^2 - 2 \vert < \epsilon$, for a small $\epsilon = 0.05$\footnote{The sample variance on $\chi^2$ (at each epoch, assuming a $\chi^2$-distribution) for the validation set can be expected to be $\sigma^2 = 2N_\mathrm{par}/N_\mathrm{validation} \sim (0.003)^2$.}. %
        We use the Adam optimizer \citep{Adam_2014arXiv1412.6980K} with a learning-rate of $5.8\times10^{-4}$. The Adam moment parameters have the values $\beta_1 = 0.97$ and $\beta_2 = 0.999$. %
        We performed a Bayesian hyperparameter tuning for fixing the values of kernel weight decays, the dropout rate, the learning rate, and the optimizer moment $\beta_1$ parameter. Please refer to Appendix \ref{app:hypertuning} for a further description of our strategy for choosing optimal hyperparameters for our network architecture and training. %
        We present the progress of the network's training quantified by the four metrics mentioned above in Appendix~\ref{app:learning-curves}.

    \subsection{Ensemble Learning}\label{sub:ensemble}
        The initialization of our network weights (kernels) as well as the training over batches and epochs is a stochastic process. This introduces a bias in the network predictions that can be traded for variance in a set of randomly initialized and trained networks. Essentially, if the errors in different networks' predictions are uncorrelated, then combining the predictions of multiple such networks helps in improving the accuracy of the predictions. It has been shown that a ``committee'' of neural networks could outperform even the best-performing member network \citep{Dietterich_Ensemble_Learning}. This falls under the umbrella of ``ensemble learning''.

        Once we have an optimal set of hyperparameter values for {\sansa}, we train 20 neural networks with the exact same architecture and the learning hyperparameters but initializing the network weights with different pseudo-random seeds and training with differently shuffled and augmented batches of the data-set. The output predictions by all the member networks of this committee of $N_{\sansa} = 20$ neural networks are then combined in the form of a weighted averaging  of the individual predictions to obtain the final outcome (this is commonly known as bootstrap aggregating or ``bagging''; see e.g. \citealt{Breiman_Bagging}). For a given input spectrum $x$, let $\mathcal{S}_i(x)$ denote the output point predictions by the $i$th network in our committee. Then the combined prediction of the committee is
        \begin{equation}\label{eqn:committee-combination}
            \mathcal{S}(x) \propto \frac{1}{N_{\sansa}} \sum_i \frac{1}{\vert \mathbf{\tilde{C}}_i(x) \vert} \mathcal{S}_i (x),
        \end{equation}
        where $\mathbf{\tilde{C}}_i (x)$ is the output estimate of the covariance matrix by the $i$th network for the input spectrum $x$. This combination puts more weight on less uncertain network predictions and thus is optimally informed by the individual network uncertainties. Even with such a small number of cognate members, we observe slight improvements \emph{w.r.t.} the best-performing member as discussed in Appendix~\ref{app:single-vs-committee}. All the output point predictions by {\sansa} considered in the following part of the text are implicitly assumed to be that of the committee and not of an individual network unless specified otherwise. %

    \subsection{Inference}\label{sub:lyanna-inference}
        \begin{figure}
            \centering
            \includegraphics[width = \linewidth]{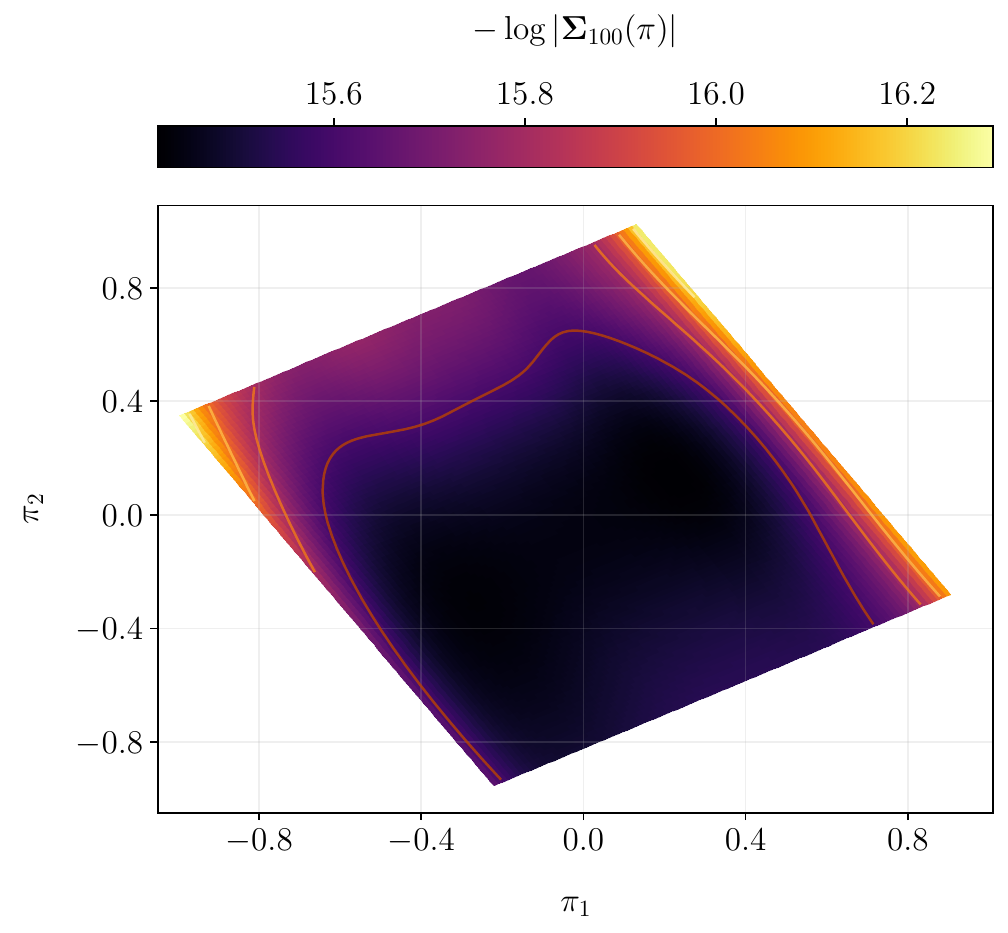}
            \includegraphics[width = \linewidth]{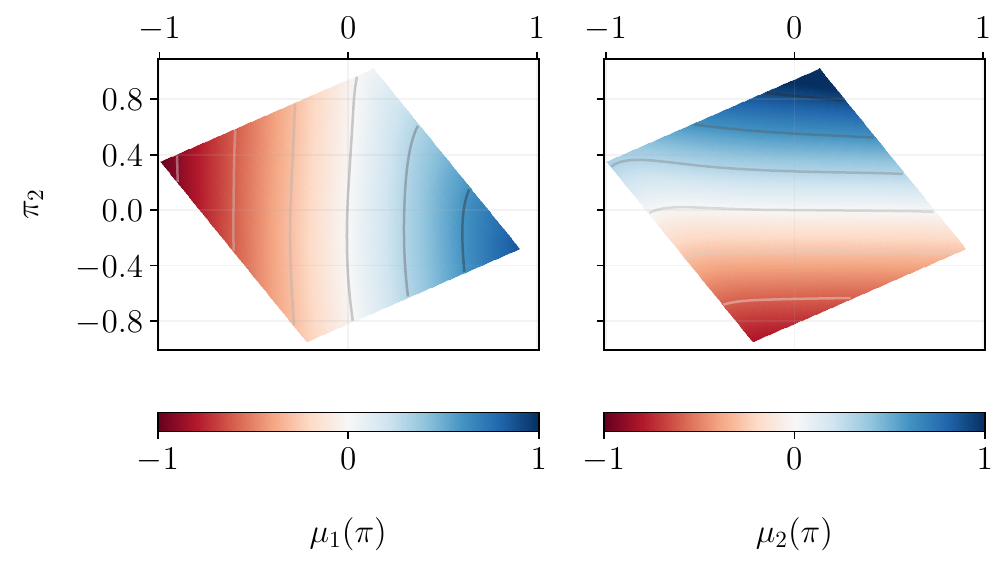}
            \caption{The likelihood model of inference with {\sansa} over our prior range for the rescaled $\pi$ parameters. The top panel shows a measure of the covariance model $\mathbf{\Sigma(\pi)}$ and the bottom panels the model for the two rescaled parameters, $\mu_1 \equiv \tilde{T}_0$ and $\mu_2 \equiv \tilde{\gamma}$. The model for the mean summary vector fairly approximates $\mu_i(\pi) = \pi_i$. 
            }
            \label{fig:likelihood-cov-model}
        \end{figure}
        We perform Bayesian inference of the model parameters with {\sansa} as well as the traditional summary statistics introduced in Section~\ref{sub:summaries}. In all the cases, we assume a Gaussian likelihood and a uniform prior over the range shown in \autoref{fig:orthogrid_space}. For inference with {\sansa} we create an emulator for a likelihood analysis in the following way. A test set of spectra\footnote{Note that this is the same set of spectra as that used for the validation of generalization during network training (see Section~\ref{sub:training}).} for a given truth $(\hat{T}_0, \hat{\gamma})$ are fed into {\sansa} and a corresponding set of parameter point estimates $(\tilde{T}_0, \tilde{\gamma})$ are obtained. Owing to our optimization strategy (described in Section~\ref{sub:training}) %
        these network predictions have an inherent scatter that is consistent with a network covariance estimate $\mathbf{\tilde{C}}$. A mean point prediction $\bar{\pi}=(\bar{T}_0, \bar{\gamma})$ and a covariance matrix is estimated from the scatter of the point estimates. %
        This is performed for each of our 121 thermal models in the test sample. %
        We then cubic-spline %
        interpolate the mean network point prediction and the %
        scatter covariance\footnote{For computational simplicity, we actually interpolate the inverse of the scatter covariance matrix}  over our prior range of thermal parameters $\mathbf{\pi}$ to obtain a model (emulator)%
        $\big[ \mathbf{\mu(\pi)}, \mathbf{\Sigma(\pi)} \big]$. The advantage of creating such a model for the likelihood is two-fold: (i) we can perform Bayesian inference with a different choice of prior (e.g. Gaussian) within the grey-shaded area of \autoref{fig:orthogrid_space}, and (ii) the inference results of our machinery could, in principle, be combined with other probes of interest to further constrain our knowledge of the thermal state of IGM. This emulator is then used to perform an MCMC analysis for getting posterior constraints with a likelihood function, %
        \begin{multline}
            \log L_N(\mathbf{\pi})\\ 
            \sim -\frac{1}{2}\Big[\log \vert \mathbf{\Sigma}_N(\pi) \vert - (\bar{\pi}_N - \mu(\pi)) \mathbf{\Sigma}_N\mathbf{^{-1}(\pi)} (\bar{\pi}_N - \mu(\pi))^T\Big],
        \end{multline}\label{eqn:sansa-likelihood-model}
        where $\bar{\pi}_N$ is the mean network point prediction for a given set of $N$ test spectra and $\mathbf{\Sigma}_N (\pi)= \mathbf{\Sigma} (\pi)/N$ quantifies the uncertainty in the mean point estimate for the given data-set size\footnote{Formulated thus (and due to the Gaussian likelihood assumption), $N$ could be deliberately varied to mimic the inference outcome of a given size of the data-set.}. We show this model in \autoref{fig:likelihood-cov-model}. The model for the mean parameter values, $\mu(\pi)$, consists of rather smoothly varying functions approximating $\mu_{1,2}(\pi_1, \pi_2) = \pi_{1,2}$, conforming to our expectation.

\section{Results and Discussion} \label{sec:results}

    In this section, we show the results of doing mock inference with our machinery as described in Section~\ref{sec:neural-nets} %
    and compare them with a summary-based approach (see Section~\ref{sub:summaries} for more details on the summaries used). We investigate a few different test scenarios for establishing robustness of our inference pipeline. For each test case, we quantify our results in two chief metrics:\\ 
    (i) precision, in terms of the area of posterior contours as a figure of merit (FoM), 
    \begin{equation}
        \mathrm{FoM} \sim 1/\sqrt{\vert \mathbf{C} \vert};
    \end{equation}\\ 
    and (ii) accuracy, in terms of a reduced $\chi^2$,
    \begin{equation}
         \delta \chi^2_\mathrm{r} \equiv \langle \mathbf{\Delta C^{-1}} \mathbf{\Delta}^T \rangle / N_\mathrm{par} - 1;
    \end{equation}\\
    where $\mathbf{\Delta}_i = (\pi_i - \hat{\pi})$ is a point in the posterior MCMC sample, $\mathbf{C}$ is a covariance matrix of $\pi$ estimated from the posterior sample and the average $\langle \rangle$ is taken over the entire sample. Note that in the two-parameter case, the area of the posterior contours is proportional to $\sqrt{\vert \mathbf{C} \vert}$. We expect that the FoM improves when including more information (about the parameters of interest) in the underlying summary statistic from the transmission field, since the constraints get tighter (contours smaller) as a consequence. A smaller value of $\delta \chi^2_\mathrm{r}$ implies a more accurate recovery of the true parameters.

\begin{figure}
        \centering
        \includegraphics[width = 0.9\linewidth]{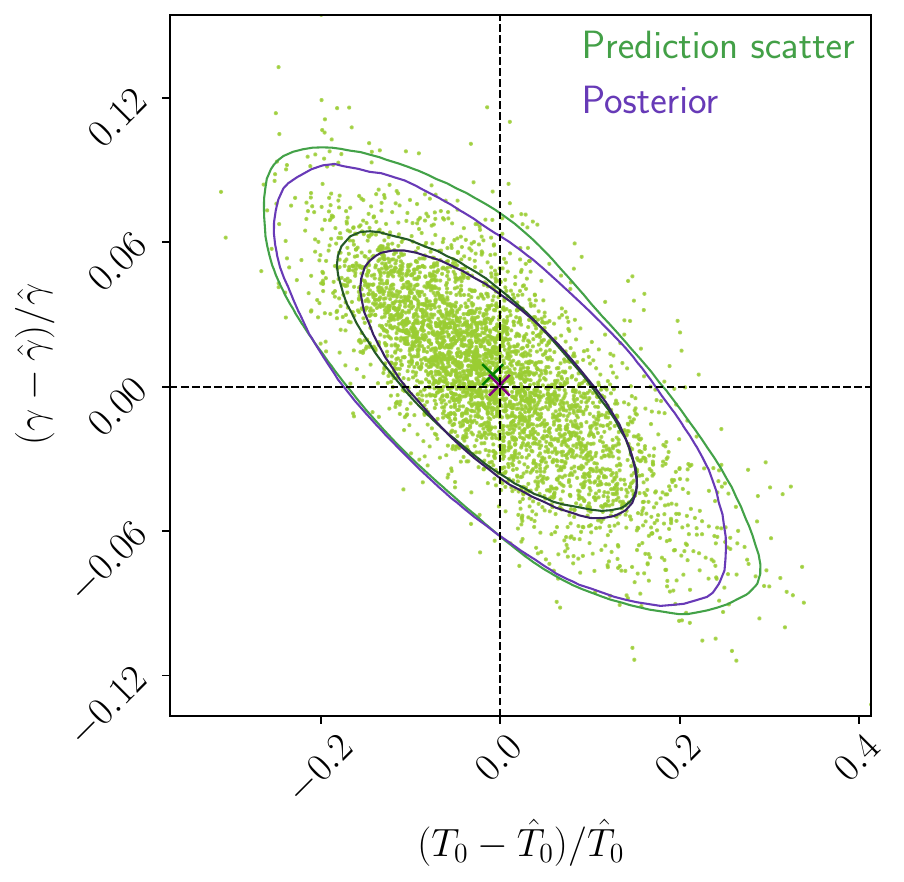}
        \caption{Scatter of %
        point predictions for the original test set of spectra from our fiducial TDR model shown here with the 68\% and 95\% contours (green). The contours of the posterior distribution (purple) obtained by {\sansa} with the procedure outlined in Section~\ref{sub:lyanna-overview}, inflated to match the information equivalent to one spectrum, follow the scatter contours very closely. {\sansa} also recovers the true parameters (dashed) with a very good accuracy, as indicated by the mean of the point prediction scatter (green cross) as well as that of the posterior (purple cross). %
        }
        \label{fig:scatter_gaussian_posterior_contours}
    \end{figure}

\begin{figure}
        \centering
        \includegraphics[width=\linewidth]{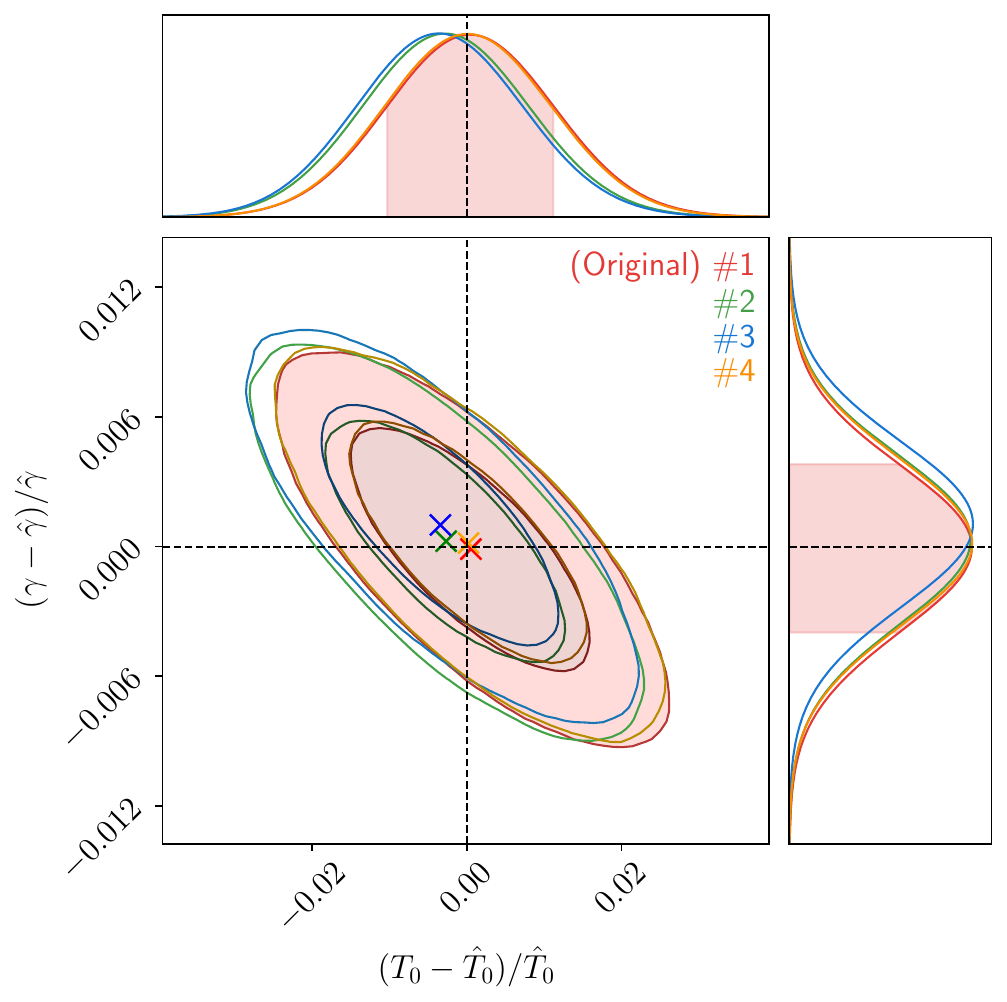}
        \caption{A comparison of posterior contours obtained for four different sets of 4,000 spectra, where \#1 is the ``original'' (for information equivalent to 100 spectra).}
        \label{fig:contours_three_sets}
    \end{figure}

    First, we consider a test set of spectra that are distinct from those used in training and validation to evaluate the performance of our inference pipeline for previously unseen data (recall that we build the likelihood model for {\sansa} using the validation set). This set consists of 4,000 spectra for the underlying true (fiducial) thermal model, $\hat{T}_0 = 10104.15$ K and $\hat{\gamma} = 1.58$ (note that this model is off the training grid, as shown in \autoref{fig:orthogrid_space}, affording us a test of the machinery's performance off-grid). To distinguish this set with the other test sets in the following, we call it the ``original'' set hereinafter. In \autoref{fig:scatter_gaussian_posterior_contours}, we show the output scatter of point estimates by {\sansa} for the original test set, with contours of 68\% and 95\% probability. For comparison, we also plot %
    the posterior contours (obtained by {\sansa} following the strategy outlined in Section~\ref{sub:lyanna-overview}), inflated to emulate the uncertainty equivalent to one input spectrum. A very good agreement 
    is observed %
    between both the cases, suggesting that %
    a cubic-spline interpolation of the scatter covariance %
    is a sufficiently good emulator for a likelihood analysis %
    as discussed in Section~\ref{sub:lyanna-inference}. %
    We perform inference with three further previously-unseen sets of 4,000 random spectra with the fiducial TDR to establish the statistical sanity of our pipeline. We show the posterior contours obtained for those in \autoref{fig:contours_three_sets} and the metric values in \autoref{tab:metrics_three_sets}.
    \begin{table}
        \caption{Comparison metric values for {\sansa} with four distinct sets of test spectra (for information equivalent to 100 spectra)}
        \label{tab:metrics_three_sets}
        \centering
        \begin{tabular}{l r r}
             \hline
             Test set & $\delta \chi^2_\mathrm{r}$ & FoM / FoM(orig.)\vspace{4pt}\\
             \hline
             \#1 (orig.) & 0.002 & - -\\
             \#2 & 0.053 & 0.994 \\
             \#3 & 0.059 & 1.004\\
             \#4 & 0.004 & 1.003\\
             \hline
        \end{tabular}
    \end{table}
    
\begin{figure}
        \centering
        \includegraphics[width=\linewidth]{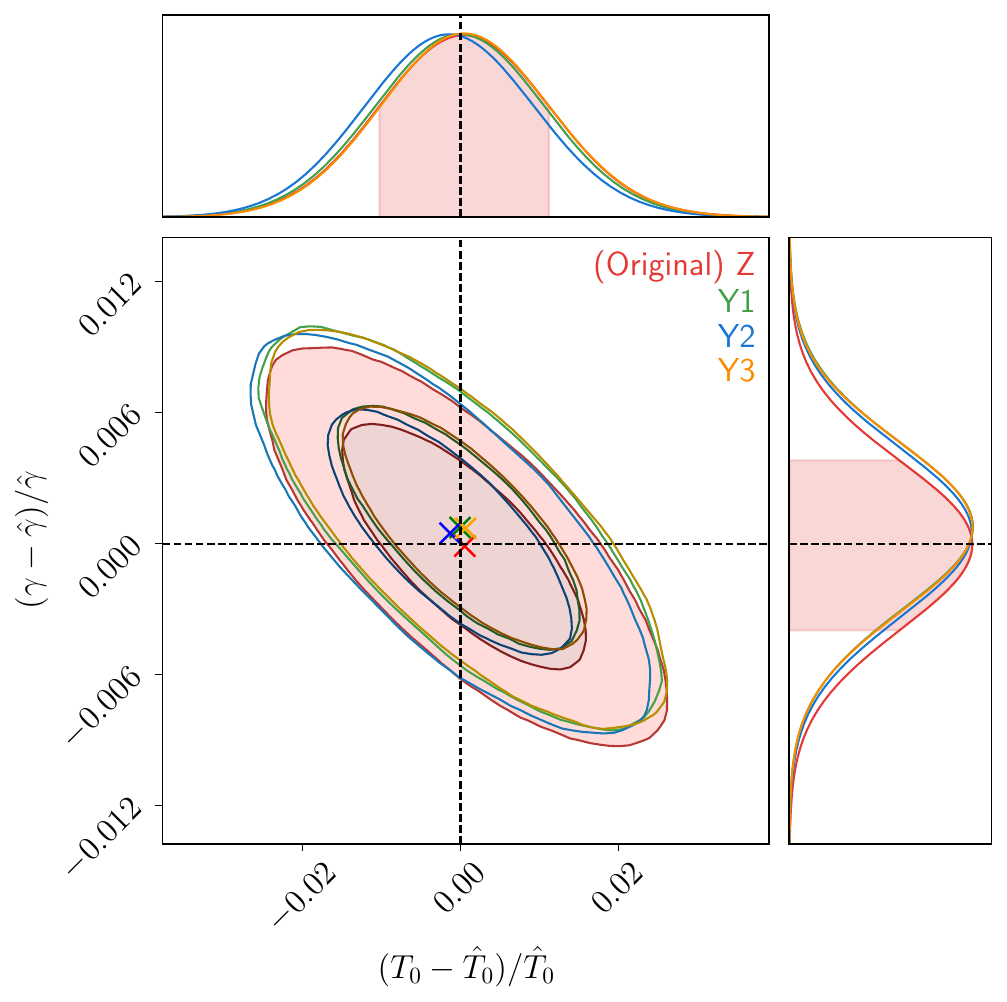}
        \caption{A comparison of posterior contours obtained for three different sets of 4,000 spectra computed along another axis (``Y'') of the simulation box (for information equivalent to 100 spectra). ``Z'' corresponds to the ``original''.}
        \label{fig:contours_y_extracted}
    \end{figure}

    Skewers can be picked along any of the three axes of the simulation box (see Section~\ref{sub:mocks}), each leading to a different realization of the {\lya} transmission, mimicking cosmic variance. We expect our pipeline to be robust to the choice of axis along which the input skewers are extracted. We choose three different test sets of skewers along another axis (``Y'') of our box that have the same underlying thermal parameters (fiducial). We estimate the posterior constraints for all three (Y1,2,3) data-sets with {\sansa} and show them in \autoref{fig:contours_y_extracted}, along with the ``original'' (skewers extracted along the Z-axis of the simulation box). The corresponding metric values are listed in \autoref{tab:metrics_y_extracted}. We observe a statistically consistent posterior distribution in each of the three Y-extracted test cases with the original case, indicating that {\sansa} is agnostic to the choice of LOS direction, even though it is trained only with one of the three possibilities. 
    \begin{table}
        \caption{Comparison metric values for {\sansa} with spectra along different axes of the simulation (for information equivalent to 100 spectra)}
        \label{tab:metrics_y_extracted}
        \centering
        \begin{tabular}{l r r}
             \hline
             Test set & $\delta \chi^2_\mathrm{r}$ & FoM / FoM(orig.)\vspace{4pt}\\
             \hline
             Original & 0.002 & - - \\
             Y1 & 0.043 & 0.998 \\
             Y2 & 0.010 & 0.997\\
             Y3 & 0.064 & 1.003\\
             \hline
        \end{tabular}
    \end{table}

\begin{figure}
        \centering
        \includegraphics[width=\linewidth]{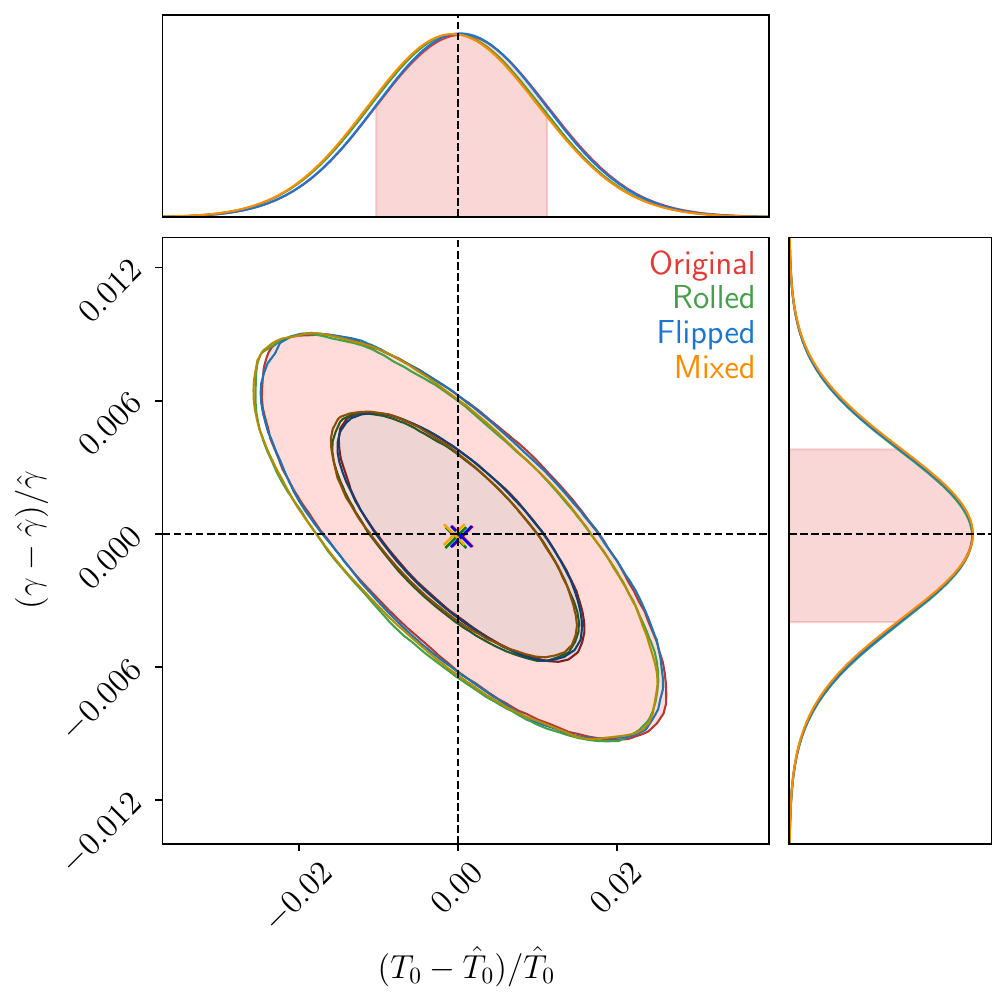}
        \caption{A comparison of posterior contours for differently augmented test spectra. In the ``rolled'' case, a uniform random amount (between 1 and 512 pixels) of cyclic permutation of the pixels is applied to each spectrum in the original test set. An arbitrary 50\% of spectra from the original test set are flipped (mirrored) in the ``flipped'' case. A random mix of both is applied in the ``mixed'' case. All of the contours carry information equivalent to 100 spectra. A mean (expectation) value of all the posterior distributions is also shown with a cross of the corresponding color. The posterior contours for all the cases agree extremely well with the original test case.}
        \label{fig:contours_augmentation}
    \end{figure}

    Moreover, we test our inference machinery with numerically modified (augmented) spectra, as discussed in Section~\ref{sub:training}. In one case, we apply a cyclic permutation of the pixels (rolling) by a random amount (between 1 and 512 pixels) to each spectrum of our original test set. We denote this by ``rolled''. In a second, we flip an arbitrary 50$\%$ of the original test set of spectra, denoted by ``flipped''. We generate another set of spectra with a random mix of both of the above operations applied to the original set; this is labeled ``mixed''. We present the posterior constraints for all of these cases in \autoref{fig:contours_augmentation} and list the metric values in \autoref{tab:metrics_augmentation}. The posterior constraints in all the augmented test scenarios agree very well with each other and with the original test case, establishing robustness of the inference against such degeneracies.
    \begin{table}
        \caption{Comparison metric values in data-augmentation scenarios for {\sansa} (for information equivalent to 100 spectra)}
        \label{tab:metrics_augmentation}
        \centering
        \begin{tabular}{l r r}
             \hline
             Test set & $\delta \chi^2_\mathrm{r}$ & FoM / FoM(orig.)\vspace{4pt}\\
             \hline
             Original & 0.002 & - -  \\
             Rolled & 0.005 & 0.997 \\
             Flipped & 0.001 & 1.001\\
             Mixed & 0.003 & 1.003 \\
             \hline
        \end{tabular}
    \end{table}
    
    \begin{figure*}
        \centering
        \includegraphics[width=0.75\linewidth]{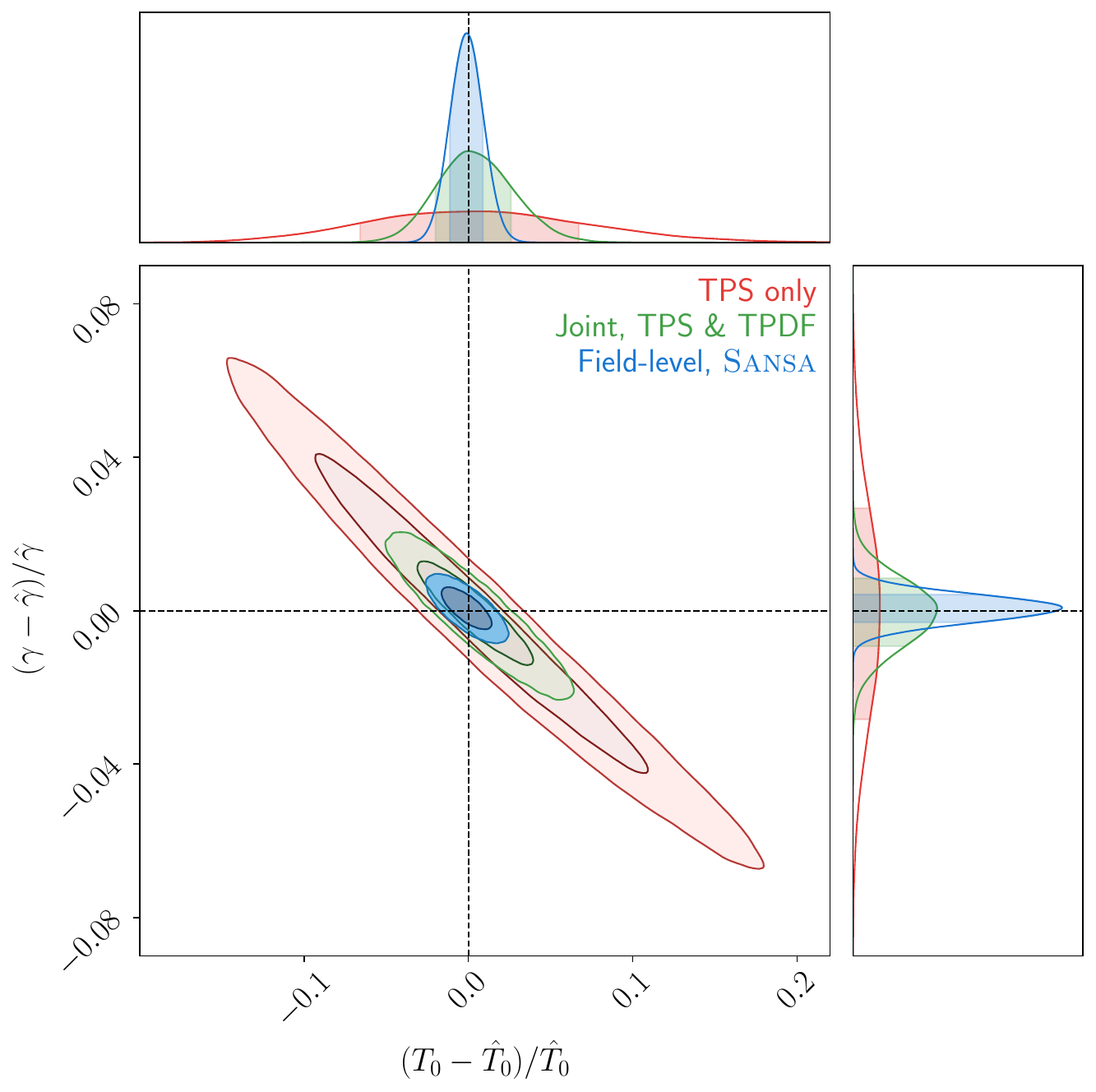}%
        \caption{Posterior contours obtained by {\sansa} for the underlying fiducial thermal model. The posterior contours from the two traditional summary statistics are shown for comparison: (i) TPS only and (ii) joint constraints of TPS and TPDF where the cross-correlations of the summaries are accounted for by a joint covariance matrix. In terms of the size of the contours, the {\sansa} constraints are tighter than the TPS-only ones by a factor 10.92 and the joint constraints by 3.30, corroborating the claim that {\sansa} recovers relevant information for inference that is not carried by the TPS and/or the TPDF. (All the cases carry information equivalent to 100 spectra.)}
        \label{fig:contours-mcmc-sansa}
    \end{figure*}
    Finally, we compare the inference outcome of {\sansa} with the traditional summary statistics (TPS and TPDF) based procedure. We present the posterior constraints on $(\pi - \hat{\pi})/\hat{\pi}$ obtained by a MCMC analysis of TPS only, TPS and TPDF jointly, and {\sansa} for the fiducial thermal model in \autoref{fig:contours-mcmc-sansa}. Evidently, the joint constraints of the two summary statistics are tighter than the TPS-only case as there is more information of the thermal parameter in the former. However, by far the field-level constraints by {\sansa} are tighter than both the traditional summary statistics cases, namely, a factor of 10.92 compared to the TPS-only case and a factor of 3.30 compared to the joint constraints in our FoM. Indeed, the TPS is only a two-point statistic of the transmission field that has a highly non-Gaussian one-point PDF itself. Combining TPS and TPDF provides some more leverage, however, it still fails to account for some relevant parts of the information for inference. As illustrated by \autoref{fig:contours-mcmc-sansa}, {\sansa} provides a remedy to the lost information by trying to optimally extract all the features of relevance at field-level.

\section{Conclusion}\label{sec:conclusion}
    We built a convolutional neural network called {\sansa} for inference of the thermal parameters $(T_0, \gamma)$ of the IGM with the {\lya} forest at field-level. We trained this using a large set of mock spectra extracted from the {\Nyx} hydrodynamic simulations. For estimating posterior constraints, we created a reasonably robust pipeline that relies on the point predictions of the parameters and the uncertainty estimates by our neural network and that can in principle be easily combined with multiple other probes of the thermal state of the IGM. A comparison of our results with those of traditional summary statistics (TPS and TPDF in particular) revealed an improvement of posterior constraints in area of the credible regions by a factor 10.92 \emph{w.r.t.} TPS-only and 3.30 \emph{w.r.t.} a joint analysis of TPS and TPDF. We established statistical robustness of our pipeline by performing tests with a few different sets of input spectra.
    
    However, our neural network that is trained with noiseless mock spectra for inference fails for spectra with even very small noise (as small as having a continuum-to-noise ratio of 500 per 6 km/s). Indeed, our framework must be adapted for use with noisy spectra by retraining {\sansa} with data-sets containing artificially added noise that varies on-the-fly during training (to prevent learning from the noise). 
    
    Furthermore, in this work we have assumed a fixed underlying cosmology for generating the various data-sets used for training and inference. However, the {\lya} forest carries information about the cosmological parameters that may be correlated with the thermal properties, and as a consequence, our machinery would exhibit a bias if the cosmology of the training data were not equal to that of the test case. A generalized pipeline would thus require marginalization over the cosmological parameters. Our proof-of-concept analysis, however, opens up an avenue for constraining cosmological parameters at field-level as well.

    Baryonic feedback from AGN and supernovae (and similarly inhomogeneous reionization) affects the phase-space distribution of IGM and is thus expected to influence the performance of our neural machinery. Nonetheless, we anticipate this impact to be marginal since the network is more sensitive to the power-law regime of the diffuse gas which still holds to a large degree.

    As described in Section~\ref{sub:mocks}, we used snapshots of the skewers to train our pipeline instead of accounting for a lightcone evolution of the IGM properties along the LOS pixel-by-pixel. We performed a test of this framework with an approximate model of such evolution; see Appendix~\ref{app:lightcone}. For the length of our skewers, we expect the actual lightcone evolution of the gas properties to be marginal and as such unproblematic for network inference.

    Nevertheless, in the spirit of creating a robust pipeline for highly realistic spectral data-sets, a plethora of physical and observational systematic effects (such as limited spectral resolution, sky-lines, metal absorption lines, continuum fitting uncertainty, damped {\lya} systems on the observational side; and lightcone effects, baryon feedback, cosmological correlations on the modeling side) must also be incorporated in the training data. This warrants a further careful investigation into training supervised deep learning inference algorithms with a variety of accurately modeled systematics added to our mock {\lya} forest data-sets and we plan to carry it out in future works.

\begin{acknowledgements}
    We thank all the members of the chair of Astrophysics, Cosmology and Artificial Intelligence (ACAI) at LMU Munich for their continued support and very interesting discussions. We acknowledge the Faculty of Physics of LMU Munich for making computational resources available for this work. %
    We acknowledge PRACE for awarding us access to Joliot-Curie at GENCI@CEA, France via proposal 2019204900. We also acknowledge support from the Excellence Cluster ORIGINS which is funded by the Deutsche Forschungsgemeinschaft (DFG, German Research Foundation) under Germany’s Excellence Strategy -- EXC-2094 -- 390783311. PN thanks the German Academic Exchange Service (DAAD) for providing a scholarship to carry out this research. 
\end{acknowledgements}

\section*{Data Availability}
The data/code pertaining to the analyses carried out in this paper shall be made available upon reasonable request to the corresponding author.

\bibliographystyle{aa}
\bibliography{main} %

\appendix

\section{An approximate lightcone model}\label{app:lightcone}
    
\begin{figure}
        \centering
        \includegraphics[width=\linewidth]{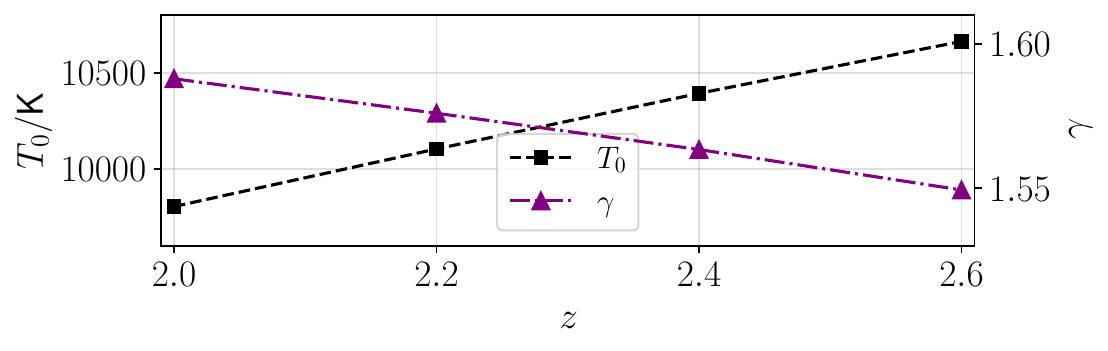}
        \caption{TDR parameters at four different redshifts; a generally smooth, linear variation can be seen in both, $T_0$ and $\gamma$.}
        \label{fig:tdr_params_z}
    \end{figure}

    Due to unavailability of lightcone simulations, our skewers entirely come from a single cosmic epoch (snapshot) and do not capture any evolution of underlying hydrodynamic fields along the LOS. Consequently the spectra carry this modeling uncertainty. However, we implemented an approximate model of lightcone evolution for testing our machinery, by interpolating certain physical quantities among close-by snapshots (in $z$). For four snapshots at $z\sim$ 2.0, 2.2, 2.4 and 2.6, we estimated the true underlying TDR parameters (see \autoref{fig:tdr_params_z}) and then linearly interpolated them to obtain $T_0(z)$ and $\gamma (z)$. The redshift span of our skewers is $\Delta z \sim 0.1$. Assuming the centers of our skewers at $z \sim 2.2$, we rescaled all the skewer temperatures according to the interpolated lightcone TDR for non-lightcone $\rho_\mathrm{b}/\bar{\rho}_\mathrm{b}$ (with the same procedure for sampling in the parameter space described in Section~\ref{sub:hydro-sims}).
     
\begin{figure}
        \centering
        \includegraphics[width=0.9\linewidth]{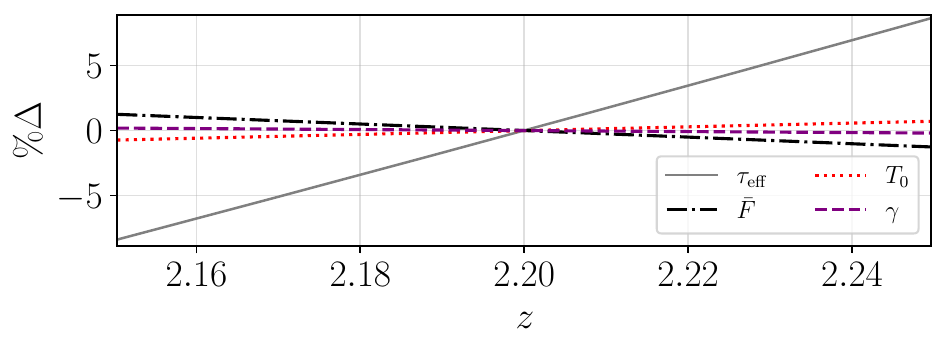}
        \caption{Percentage variations in the TDR parameters and the {\lya} transmission in our approximate lightcone model for the redshift span of our skewers.}
        \label{fig:percentage-lightcone}
    \end{figure}

    Additionally, we incorporated the overall evolution of mean {\lya} transmission through rescaling $\tau$ at pixel-level as follows. We first estimated $\tau_\mathrm{eff}(z)=-\log \bar{F}$ along our skewers from \cite{Becker_2013MNRAS.430.2067B} and then rescaled all the non-lightcone optical depth values as $\tau_0(z) \to \tau_\mathrm{lc}(z) = \tau_\mathrm{eff}(z)/\tau_\mathrm{eff}(z=2.2) \cdot \tau_0(z)$ to eventually obtain mock {\lya} transmission spectra. \autoref{fig:percentage-lightcone} shows the fractional variation in the TDR parameters and the {\lya} transmission across our skewers according to this approximate lightcone model. A maximum of $\lesssim 1.5 \%$ deviation in $T_0$ and $\lesssim 0.4 \%$ in $\gamma$ can be seen. In $\bar{F}$, there is a maximum of $\lesssim 2.6 \%$ variation across the redshift span of the skewers. We followed the above procedure for generating 4,000 lightcone spectra with the same 4,000 physical skewers as the ``original'' test case (Section~\ref{sec:results}).
    
    \begin{figure}
        \centering
        \includegraphics[width=\linewidth]{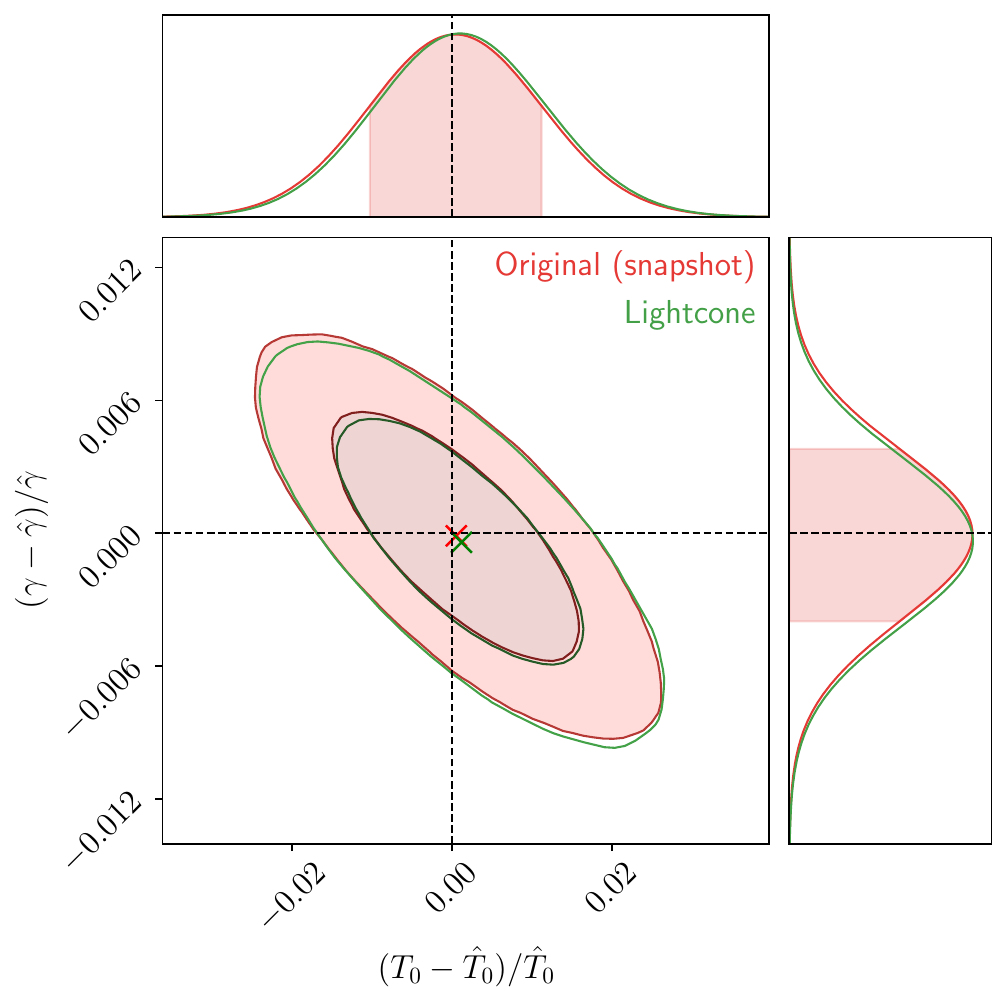}
        \caption{A comparison of posterior contours between the original (snapshot) and an approximate lightcone model test cases. Both carry information equivalent to 100 spectra. A mean (expectation) value of the posterior distributions are also shown with crosses of the corresponding colors. In both the cases, statistically inter-consistent posterior distributions are obtained, recovering the fiducial TDR of the snapshot.}
        \label{fig:contours-lightcone}
    \end{figure}
    
    With these 4,000 lightcone spectra, we performed inference with {\sansa}; see the posterior constraints in \autoref{fig:contours-lightcone}. We expect {\sansa} to be able to recover a ``mean'' TDR along the skewers, i.e., the thermal parameters at the centres of the skewers, $\hat{T}_0 = 10104.15$ K 
    and $\hat{\gamma} = 1.58$ (the fiducial values). The metrics for the lightcone and the original test cases are compared in \autoref{tab:metrics_lightcone}.  
    \begin{table}
        \caption{Comparison metric values for the original (snapshot) and an approximate lightcone test cases for {\sansa} (for information equivalent to 100 spectra)}
        \label{tab:metrics_lightcone}
        \centering
        \begin{tabular}{l r r}
             \hline
             Test set & $\delta \chi^2_\mathrm{r}$ & FoM / FoM(orig.)\vspace{4pt}\\
             \hline
             Snapshot (orig.) & 0.002 & - - \\
             Lightcone & 0.008 & 1.001 \\
             \hline
        \end{tabular}
    \end{table}

\section{Orthogonal basis of the parameters}\label{app:orthogrid}

        Heuristically, the training of the network is most efficient when our training sample captures the most characteristic variations in the data \emph{w.r.t.} the two parameters of interest, $T_0$ and $\gamma$. Indeed, as found by many previous analyses (e.g., \citealt{Walther_IGM_2019ApJ...872...13W}), there appears to be an axis of degeneracy in the said parameter space given by the orientation of the elongated posterior contours. This presents us with an alternative parametrization of the space accessible via an orthogonalization of a parameter covariance matrix. By doing a mock likelihood analysis with a linear-interpolation emulator of the TPS on a pre-existing grid of thermal models, we first obtain a (rescaled) parameter covariance matrix $\mathbf{C}$ and then diagonalize that such that $\mathbf{\Lambda = V}^T\mathbf{CV}$ is a $2\times 2$ diagonal matrix ($\mathbf{V}$ symmetric). An ``orthogonal'' representation of the parameters can then be found by a change of basis,
        \begin{equation}\label{eqn:change-of-basis}
            \mathbf{A}^T = \mathbf{V\pi}^T,    
        \end{equation}
        where $\mathbf{A} = (\alpha, \beta)$ and $\mathbf{\pi} = (T_0, \gamma)$. In the above definition, $\beta$ represents the degeneracy direction in the $\pi$ parameters and $\alpha$ corresponds to the axis of the most characteristic deviation. Thence, we sample the parameter space for training with an 11$\times$11 regular grid in the orthogonal parameter space.

\section{Biases due to a limited prior range}\label{app:biases}

        \begin{figure}
            \centering
            \includegraphics[width = \linewidth]{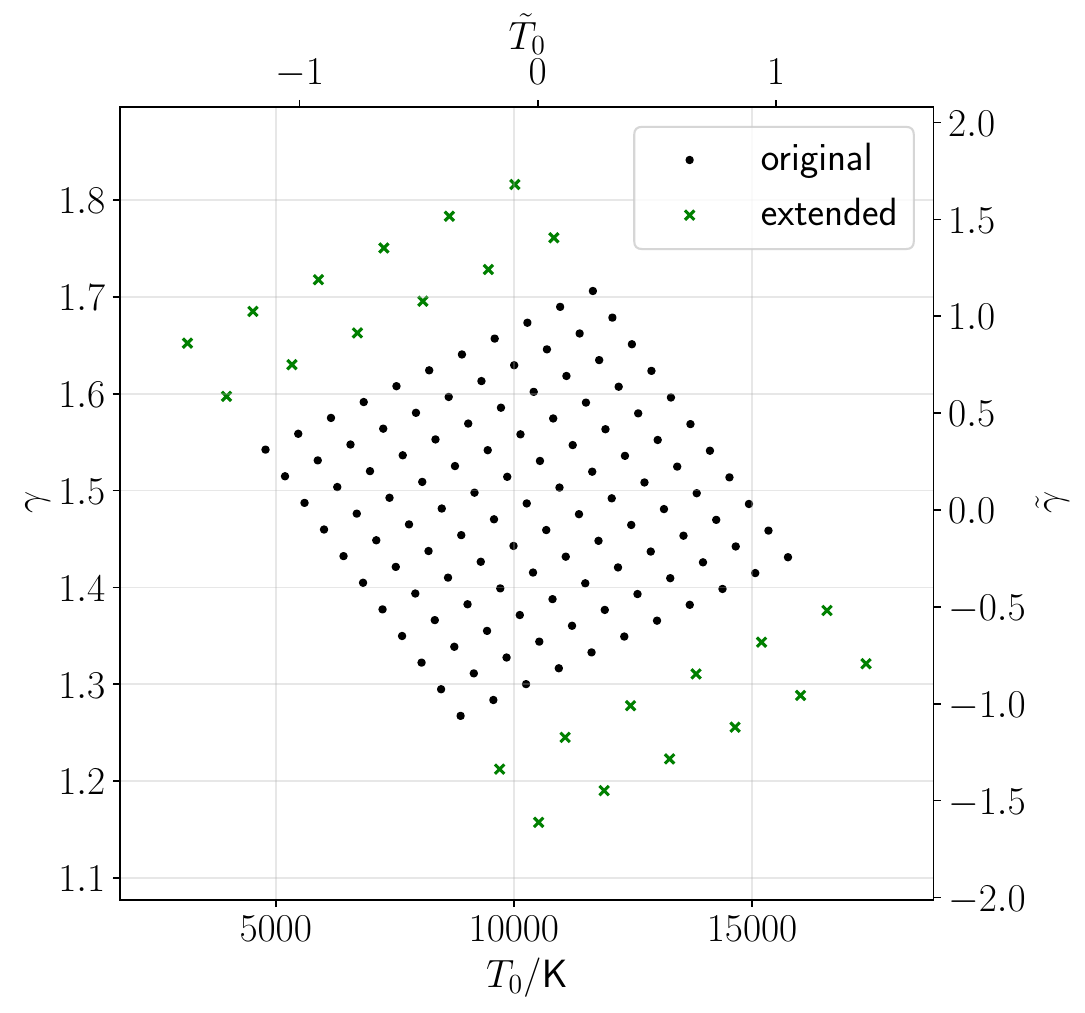}
            \caption{An extended grid of thermal models for regularizing the point predictions of {\sansa}. The extension is sparser than the actual training grid as it is only used for rectifying the skewed network predictions for models on the original grid by implicitly allowing a larger prior volume during training. No further test (or network validation) is performed for the models on the extended grid.}
            \label{fig:extorthogrid_space}
        \end{figure}
\begin{figure}
            \centering
            \includegraphics[width =\linewidth]{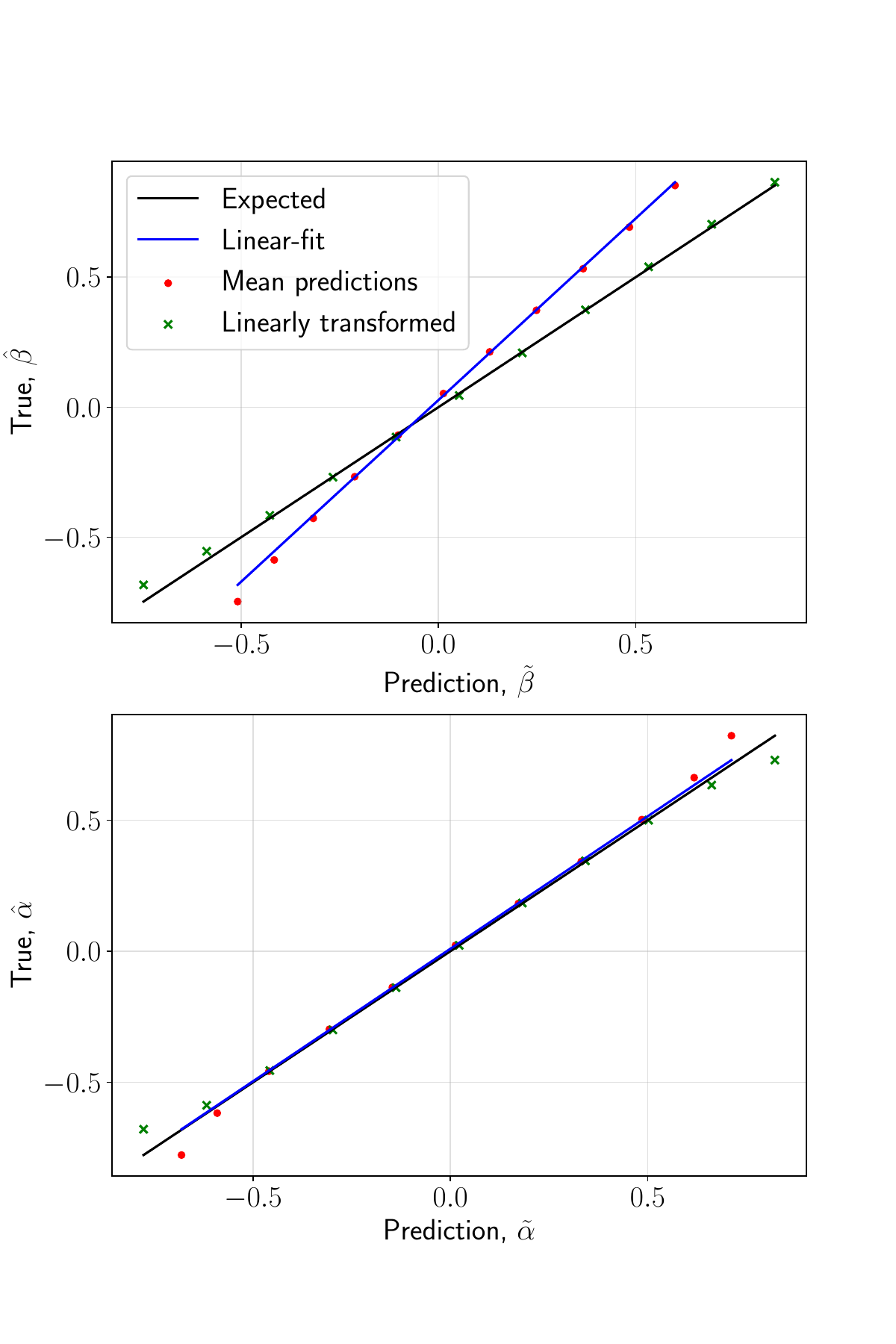}
            \caption{Mean point predictions on the validation set from a network trained on an extended grid as shown in \autoref{fig:extorthogrid_space} and a straight line fit of each independent orthogonal parameter (averaged over the other). The mean predictions after applying the linear transformation can be seen to follow the expected $y = x$ behavior.}
            \label{fig:linear_fits_alpha_beta}
        \end{figure}

        After a cursory network training we observed that when the true labels fall close to the edges of our prior range set by the training sample as shown in \autoref{fig:orthogrid_space}, the mean network predictions are biased %
        toward the center of that space. We also found that in the central region of our prior space, all the network predictions are roughly gaussian-distributed but closer to the edges of that prior they are more skewed, i.e., the mode of the distribution is closer to the ground truth than the median. To regularize this, we sampled an extended, sparser grid of thermal models along the degeneracy direction $\beta$ (in which {\sansa} provides weaker constraints), to augment our training data-set as shown in \autoref{fig:extorthogrid_space}. After retraining {\sansa} on this extended grid, we observed that a (quantifiable) bias still exists (\autoref{fig:linear_fits_alpha_beta}) but the predictions have mostly gaussianized. The mean point predictions for each of the thermal models on the original grid along a given orthogonal parameter axis fall on approximately a straight line, and hence we can perform a linear transformation of all the raw network predictions such that they satisfy our expectation, $\tilde{\pi} = \hat{\pi}$. This tractable transformation can be represented as follows. In the orthogonal parameters,
        \begin{equation}
            \mathbf{\tilde{A}}^T_{\text{f}} = \mathbf{M\tilde{A}}^T_{\text{i}} + \mathbf{c}^T, 
        \end{equation}
        where $\mathbf{\tilde{A}} = (\tilde{\alpha}, \tilde{\beta})$ is a point prediction vector in the orthogonal basis, $\mathbf{M}$ is a diagonal matrix and the subscripts ``i'' and ``f'' denote the original and transformed states of the vector respectively. This linear transformation is also shown for each parameter $\alpha$ and $\beta$ independently in \autoref{fig:linear_fits_alpha_beta}.
        Note that since the change of basis is also a linear operation, the overall transformation in $\pi$ parameters is linear and preserves the gaussianity of the point predictions. Following Eq.~\eqref{eqn:change-of-basis}, the transformation applied to the actual parameters $\tilde{\pi}$ looks like
        \begin{equation}
            \tilde{\pi}^T_\mathrm{f} = \mathbf{W} \tilde{\pi}^T_\mathrm{i} + \mathbf{d}^T,
        \end{equation}
        where $\mathbf{W = V^{-1}MV}$ and $\mathbf{d} = \mathbf{cV^{-1}}$ due to the symmetric $\mathbf{V}$. All the network covariance estimates $\mathbf{\tilde{C}}$ can also be linearly mapped with a matrix operation $\mathbf{\tilde{C}} \to \mathbf{W\tilde{C}W}^T$. We use our validation set to fit the linear transformation parameters, $\mathbf{W}$ and $\mathbf{d}$, for each trained neural network in the committee (the two parameter combinations (true labels) closest to the prior boundaries in each orthogonal parameter are not considered for this fitting). The full neural network {\sansa} presented in this paper has this tractable linear transformation incorporated as a final, unbiased layer. %

\section{Hyperparameter optimization} \label{app:hypertuning} 
        
        As in every deep learning implementation, our algorithm is defined by a large set of hyperparameters that must be tuned in order to arrive at the best possible location on the bias-variance trade-off. 
        Our hyperparameters include the dropout rate, the amplitude of the kernel regularization ($l_2$), number of residual blocks, number of filters and kernel-size in each residual block, weight-initialization, amount of downsampling per pooling layer, size of the batches of training data, learning rate, Adam $\beta_1$ parameter, etc. Two strategies were adopted to explore this vast hyperparameter space. Hyperparameters having a finite number of discrete possible values (e.g. architecture in the residual parts) were manually tuned with informed heuristic choices of values to try. %
        The rest of the hyperparameters having continuous spectra; namely kernel regularization, dropout, learning rate, and Adam $\beta_1$; were tuned with a Bayesian optimization algorithm --- based on tree-structured Parzen estimators --- for an informed search of the space and an economical use of the resources. This was performed using the pre-defined routines of the python package \textsc{Optuna} \citep{optuna}.

\section{Training progress}\label{app:learning-curves}
\begin{figure}
            \centering
            \includegraphics[width = \linewidth]{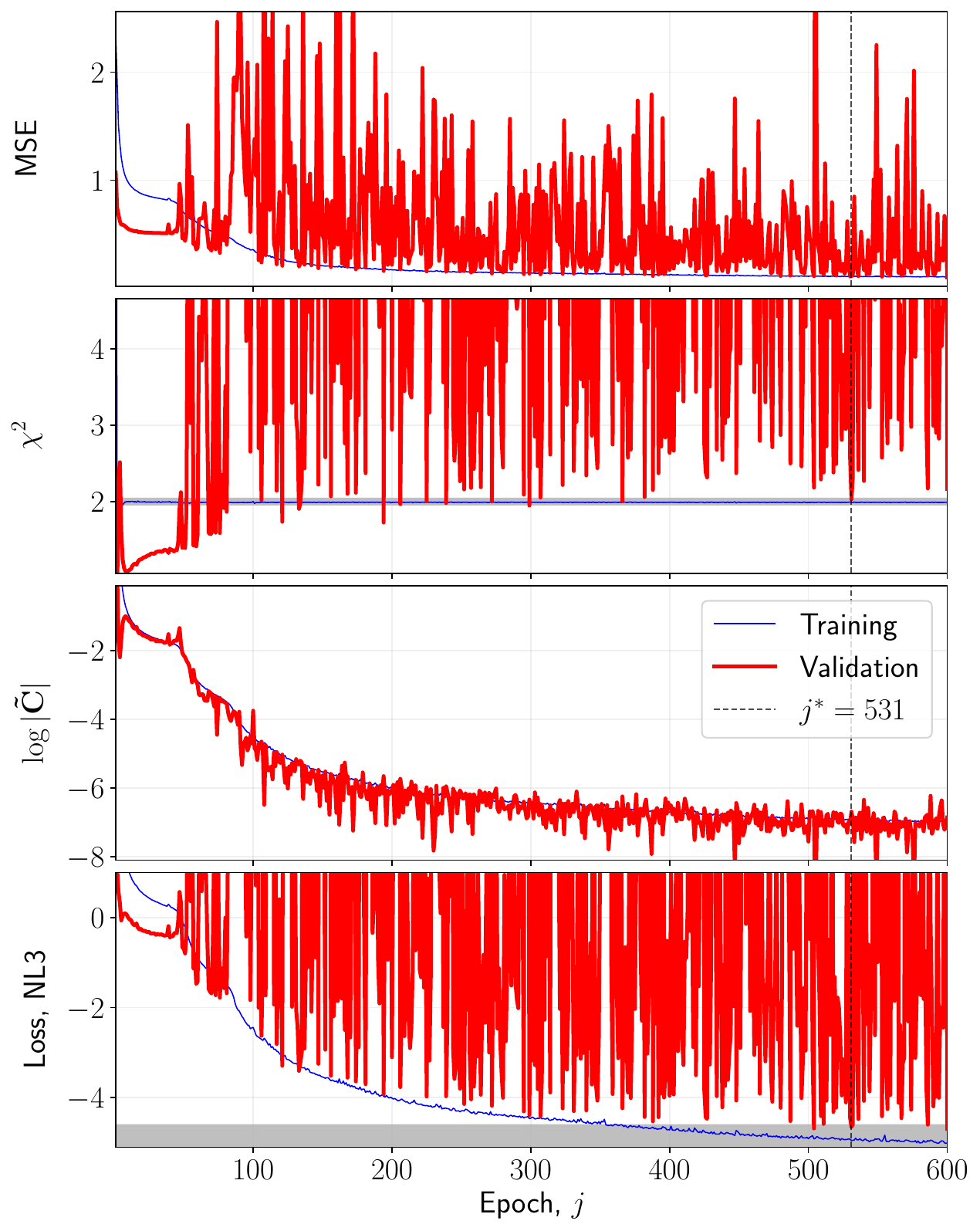}
            \caption{Learning curves of the best-performing network in the committee of {\sansa}. This network is trained for $>600$ epochs. The best state of the network that can be possibly achieved is determined by the minimal value of NL3 while $\vert \chi^2 - 2 \vert < 0.05$ for the validation set and that occurs at epoch $j^* = 531$ during training.} %
            \label{fig:sansa-learning-curves}
        \end{figure}

    We show the learning curves for {\sansa} in \autoref{fig:sansa-learning-curves} containing the four network metrics --- MSE, $\chi^2$, $\log \vert \mathbf{\tilde{C}} \vert$ and the NL3 loss --- for the best-performing network in our committee . Following our expectation, the values of the loss, MSE, and $\log \vert \mathbf{\tilde{C}} \vert$ decrease for the training set and also for the validation set (albeit somewhat stochastically) over the initial epochs and eventually the validation loss stops improving. On the other hand, $\chi^2$ for training converges to $\sim 2$ and that for validation fluctuates on the higher side, occasionally coming close to 2, with a slow overall gain over the epochs. We restore the network to its state at an epoch $j^*$ at which the loss value is minimal while $\vert \chi^2 - 2 \vert < 0.05$ for the validation set. This helps us make sure that the network predictions are generalized enough and regularized under the Gaussian likelihood cost function. For this network $j^*=531$. The same qualitative behavior is observed for all the networks in our committee. %

\section{Single network vs. committee}\label{app:single-vs-committee}
        \begin{figure}
            \centering
            \includegraphics[width=1\linewidth]{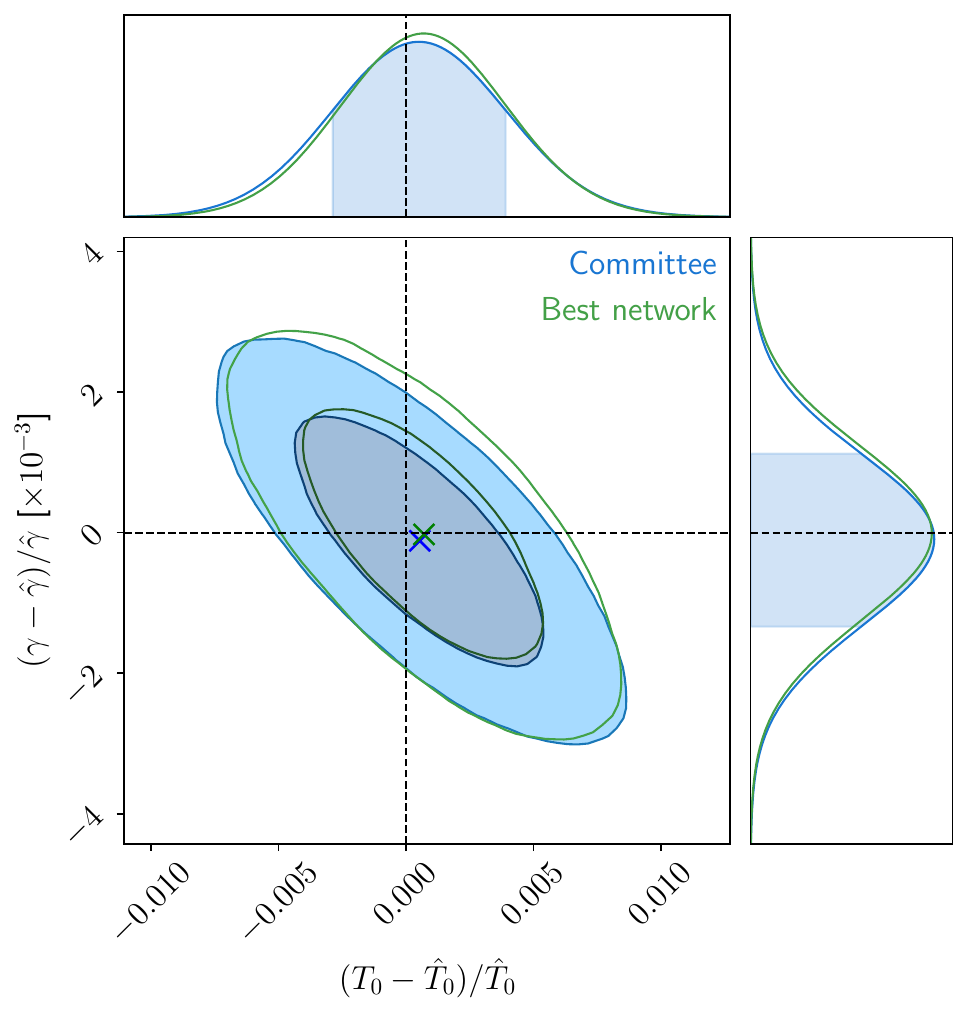}
            \caption{A comparison of the posterior contours obtained by the committee with those by the best-performing network in the committee for the fiducial thermal model, with information equivalent to 1,000 spectra. %
            The committee constraints are slightly more accurate and $\sim$4\% tighter than those of the best member network.}
            \label{fig:single-vs-committee}
        \end{figure}
        As mentioned in Section~\ref{sub:ensemble}, combining outputs of multiple similar neural networks is shown to improve the eventual outcome. Motivated by this, we employed a committee of 20 networks with the same architecture but having different initial weights, and training with different random batches and augmentation on-the-fly. In practice, the likelihood model of inference described in Section~\ref{sub:lyanna-inference} can be built for each individual network in the committee the same way as for the committee itself since the procedure relies purely on the predictions of the network(s) from the validation set. Hence, it is possible to compare the posterior constraints of an individual network with that of the committee. We present a comparison of the committee with the best-performing member network (defined as the one leading to the posterior constraints with the highest FoM of all the individual networks) in \autoref{fig:single-vs-committee} and \autoref{tab:metrics_committee_vs_best} for our ``original'' test set of spectra from the fiducial thermal model. The constraints by the committee are $\sim$ 4\% tighter than the best network in FoM and they are slightly more accurate as well.
        Even though $N_{\sansa} = 20$ is statistically a small number of sample members in a committee, the aggregate results of the ensemble are still a little better than the best-performing network, conforming to the popular findings.
        \begin{table}
            \caption{Comparison metric values of the committee and its best-performing member for our original test case for {\sansa} (for information equivalent to 1,000 spectra)}
            \label{tab:metrics_committee_vs_best}
            \centering
            \begin{tabular}{l r r}
                 \hline
                  & $\delta \chi^2_\mathrm{r}$ & FoM / FoM(committee)\vspace{4pt}\\
                 \hline
                 Committee & 0.015 & - - \\
                 Best network & 0.045 & 1.035 \\
                 \hline
            \end{tabular}
        \end{table}

\label{lastpage}
\end{document}